\documentclass[prl,twocolumn,showpacs,amsmath,amssymb,superscriptaddress]{revtex4}
\usepackage{graphicx}
\begin{document}

\title{Parameter-dependent unitary transformation approach for quantum Rabi model}
\author{Degang Zhang}
\affiliation{College of Physics and Electronic Engineering, Sichuan Normal University,
Chengdu 610101, China}
\affiliation{Institute of Solid State Physics, Sichuan Normal
University, Chengdu 610101, China}
\affiliation{Texas Center for Superconductivity and Department
of Physics, University of Houston, Houston, Texas 77204, USA}

\begin{abstract}

{\bf Abstract}:
 Quantum Rabi model has been exactly solved by employing the parameter-dependent unitary transformation method in both the occupation number representation and the Bargmann space. The analytical expressions for the complete energy spectrum consisting of two double-fold degenerate sub-energy spectra are presented in the whole range of all the physical parameters. Each energy level is determined by a parameter in the unitary transformation, which obeys a highly nonlinear equation. The corresponding eigenfunction is a convergent infinite series in terms of the physical parameters. Due to the level crossings between the neighboring eigenstates at certain physical parameter values, such the degeneracies could lead to novel physical phenomena in the two-level system with the light-matter interaction.


\end{abstract}


\maketitle

\section{I. Introduction}

The Rabi model describes the response of a two-level atom to an applied bosonic field [1]. Such a simplest interacting quantum model
has had wide applications in many fields of physics, e.g. atomic physics [2], quantum optics [3], trapped ions [4,5],
quantum dots [6], superconducting qubits [7,8,9], cold atoms [10],  and etc..
It is also expected to be the theoretical basis for quantum information and quantum technology [11-14].

Quantum Rabi model usually has the Hamiltonian
$$ H=\omega a^\dagger a+g(a^\dagger+a)\sigma_x+\lambda\sigma_z+\epsilon\sigma_x,\eqno{(1)}$$
where $\sigma_x$ and $\sigma_z$ are the Pauli matrices for the two-level system with level splitting $2\lambda$,
$a^\dagger$ and $a$ are the creation and annihilation operators for the single bosonic mode with frequency $\omega$,
respectively, the light-matter interaction is controlled by the coupling parameter $g$, and
the last term $\epsilon\sigma_x$ is the driving term which leads to tunnelling between
the two levels. We note that the competition between $g$ and $\omega$ produces the different experimental
regimes. When $g/\omega$ is small, by applying the rotating-wave approximation,
the Rabi model (1) with $\epsilon=0$ is equivalent to the so-called Jaynes-Cummings model [15],
which is relevant to most experimental regimes. Because the Jaynes-Cummings model
is integrable, it is easy to derive its analytical solution.
With increasing $g/\omega$, the ultrastrong coupling regime ($\sim 0.1<g/\omega <\sim 1.0$) [12] or
the  deep strong coupling regime ($g/\omega>\sim 1.0$) [9] is reached,
where the Jaynes-Cummings model is invalid and cannot be used to
investigate the interaction between light and matter.
Recently these regimes have rapidly growing interesting due to their
fundamental characteristics and the potential applications in quantum devices [11-14].

Although the Hamiltonian (1) has a simple form, it has not been possible to obtain its correct analytical solution,
which is considerably important for exploring accurately the light-matter interaction
from weak to extreme strong coupling. In Ref. [16],  Braak presented an analytical solution of the Rabi model (1)
by using the representation of bosonic operators in the Bargmann space of analytical functions.
The energy spectrum consists of two parts, i.e. the regular and the exceptional spectrum.
However, such a spectrum structure is incorrect due to the derivation error
in solving the time-independent Schrodinger equation in the positive and negative parity parts (see APPENDIX).

In this article, we exactly diagonalize the Hamiltonian (1) by using the parameter-dependent unitary transformation technique
in both the occupation number representation and the Bargmann space.
Such a direct and powerful approach has been used to solve successfully the complex
two-dimensional electron gas in the presence of both Rashba and Dresselhaus spin-orbit interactions
under a perpendicular magnetic field [17,18].

\section{II. Occupation number representation}

The two-component eigenstate of the Hamiltonian (1) for the nth energy level
with quantum number $s$ has the general form
$$|n,s>=\frac{1}{{\cal A}_{ns}}\sum_{m=0}^{+\infty}
\left(
\begin{array}{cc}
1&\Delta_{ns}\\
-\Delta_{ns}&1
\end{array}\right)
\left (
\begin{array}{c}
\alpha^{ns}_{m}\\
\beta^{ns}_{m}
\end{array}\right )\phi_{m},$$
$$|{\cal A}_{ns}|^2=(1+\Delta^2_{ns})\sum_{m=0}^\infty(|\alpha^{ns}_{m}|^2
+|\beta^{ns}_{m}|^2),\eqno{(2)}$$
where the $2\times 2$ matrix is a unitary one, $s=\pm 1$ are associated with the two components under the level quantum
number n, respectively, ${\cal A}_{ns}$ is the normalized factor,
$\Delta_{ns}$ is a real parameter to be determined below by requiring the coefficients $\alpha^{ns}_{m}$ and $\beta^{ns}_{m}$
to be nonzero, $\phi_{m}$ is the eigenstate of the mth energy level in the
occupation number representation, i.e. $a^+\phi_{m}=\sqrt{m+1}\phi_{m+1}$, $a \phi_{m}=\sqrt{m}\phi_{m-1}$ and $<\phi_{m^\prime}|\phi_{m}>=\delta_{mm^\prime}$. When $m \rightarrow + \infty,  \alpha^{ns}_{m}=\beta^{ns}_{m}=0$.
Substituting $|n,s>$ into the eigen-equation $H|n,s>=E_{ns}|n,s>$ and
letting the coefficients of $\phi_m$ to be zero, we obtain a coupled system of infinite homogeneous linear equations
for $\alpha^{ns}_{m}$ and $\beta^{ns}_{m}$
$$\begin{array}{rrr}
\frac{2g\Delta_{ns}}{1+\Delta_{ns}^2}(\sqrt{m}\alpha^{ns}_{m-1}+\sqrt{m+1}\alpha^{ns}_{m+1})& &\\
+[E_{ns}-m\omega-\frac{\lambda(1-\Delta^2_{ns})-2\epsilon\Delta_{ns}}{1+\Delta_{ns}^2}]\alpha^{ns}_m& &\\
-\frac{g(1-\Delta^2_{ns})}{1+\Delta_{ns}^2}(\sqrt{m}\beta^{ns}_{m-1}+\sqrt{m+1}\beta^{ns}_{m+1})& &\\
-\frac{2\lambda\Delta_{ns}+\epsilon(1-\Delta^2_{ns})}{1+\Delta_{ns}^2}\beta^{ns}_m &=&0,\\
\end{array}\eqno{(3)}$$
$$\begin{array}{rrr}
\frac{2g\Delta_{ns}}{1+\Delta_{ns}^2}(\sqrt{m}\beta^{ns}_{m-1}+\sqrt{m+1}\beta^{ns}_{m+1})& &\\
-[E_{ns}-m\omega+\frac{\lambda(1-\Delta^2_{ns})-2\epsilon\Delta_{ns}}{1+\Delta_{ns}^2}]\beta^{ns}_m& &\\
+\frac{g(1-\Delta^2_{ns})}{1+\Delta_{ns}^2}(\sqrt{m}\alpha^{ns}_{m-1}+\sqrt{m+1}\alpha^{ns}_{m+1})& &\\
+\frac{2\lambda\Delta_{ns}+\epsilon(1-\Delta^2_{ns})}{1+\Delta_{ns}^2}\alpha^{ns}_m &=&0,\\
\end{array}\eqno{(4)}$$
where $m=0, 1, 2, \cdots, \infty$, and $\alpha^{ns}_{m}=\beta^{ns}_{m}\equiv0$ for $m<0$.

\subsection{A. Sub-energy spectrum I}

In order to obtain the analytical solution of the Hamiltonian (1) in the whole parameter space,
we first choose
$$\begin{array}{rrr}
[\omega(n+1)+\frac{\lambda(1-\Delta^2_{ns})-2\epsilon\Delta_{ns}}{1+\Delta^2_{ns}}-E_{ns}]
 \alpha^{ns}_{n+1}&&\\
+\frac{g(1-\Delta^2_{ns})}{1+\Delta^2_{ns}}\sqrt{n+1}\beta^{ns}_n&=&0,
\end{array}
\eqno{(5)}$$
$$\begin{array}{rrr}
(\omega n-\frac{\lambda(1-\Delta^2_{ns})-2\epsilon\Delta_{ns}}{1+\Delta^2_{ns}}-E_{ns})
\beta^{ns}_n&&\\
+\frac{g(1-\Delta^2_{ns})}{1+\Delta^2_{ns}}\sqrt{n+1}\alpha^{ns}_{n+1}&=&0,\\
\end{array}
\eqno{(6)}$$
which come from the vanishing of the two terms about $\alpha^{ns}_{n+1}$ and $\beta^{ns}_n$
in Eq. (3) with $m=n+1$ and Eq. (4) with $m=n$, respectively.
Such a choice is based on the observation of exact solution of the
Hamiltonian (1) for the $n$th energy level with quantum number $s$ when $g=0$.
We find that the non-zero eigenfunction associated with the eigenvalue $E_{ns}$ is solely fixed by letting
$$[2\lambda\Delta_{ns}+\epsilon(1-\Delta^2_{ns})]\beta^{ns}_n
-2g\Delta_{ns}\sqrt{n+1}\alpha^{ns}_{n+1}=0,
\eqno{(7)}$$
or
$$[2\lambda\Delta_{ns}+\epsilon(1-\Delta^2_{ns})]\alpha^{ns}_{n+1}
+2g\Delta_{ns}\sqrt{n+1}\beta^{ns}_n=0.
\eqno{(8)}$$
We solve the homogenous linear equations (5) and (6) about $\alpha^{ns}_{n+1}$
and $\beta^{ns}_n$ by vanishing of the coefficient determinant.
Then the eigenvalue for the nth eigenstate with $s$ has the analytical expression
$$\begin{array}{l}
E_{ns}=(n+\frac{1}{2})\omega+s \Xi_{ns}, \\
\Xi_{ns}=\sqrt{(\frac{\omega}{2}+\frac{\lambda(1-\Delta^2_{ns})-2\epsilon\Delta_{ns}}{1+\Delta^2_{ns}}
)^2+(n+1)g^2(\frac{1-\Delta^2_{ns}}{1+\Delta^2_{ns}})^2}.
\end{array}
\eqno{(9)}$$
Note that the quasiparticle energy $E_{ns}$ must be larger than zero.
From Eqs. (6) and (7) or Eqs. (5) and (8), the parameter $\Delta_{ns}$ is determined by the highly nonlinear equation
$$\epsilon(1+\Delta_{ns}^2)-2\Delta_{ns}(E_{ns}-\omega n)=0,
\eqno{(10)}$$
or
$$\epsilon(1+\Delta_{ns}^2)+2\Delta_{ns}[E_{ns}-\omega (n+1)]=0.
\eqno{(11)}$$
After analysing carefully, we discover that Eq. (10) with $s=-1 (1)$ coincides with Eq. (11) with $s=1 (-1)$.
In other words, $\Delta_{ns}$ is independent of quantum number $s$, i.e. $\Delta_{n,1}\equiv\Delta_{n,-1}$,
which leads to $\Xi_{n,1}\equiv\Xi_{n,-1}$. So we have
$$\epsilon(1+\Delta_{ns}^2)+\Delta_{ns}(2\sigma \Xi_{ns}-\omega)=0,
\eqno{(12)}$$
where $\sigma=\pm 1$. It is easy to see from Eq. (12) that
the analytical solution (9) is physical if and only if $\Delta_{ns}\rightarrow 0$ when $\epsilon\rightarrow 0$.
Otherwise, $\Xi_{ns}\equiv\sigma\omega/2$, which is not true for arbitrary $\lambda$ and $g$.

\begin{figure}
\rotatebox[origin=c]{0}{\includegraphics[angle=0,
          height=1.24in]{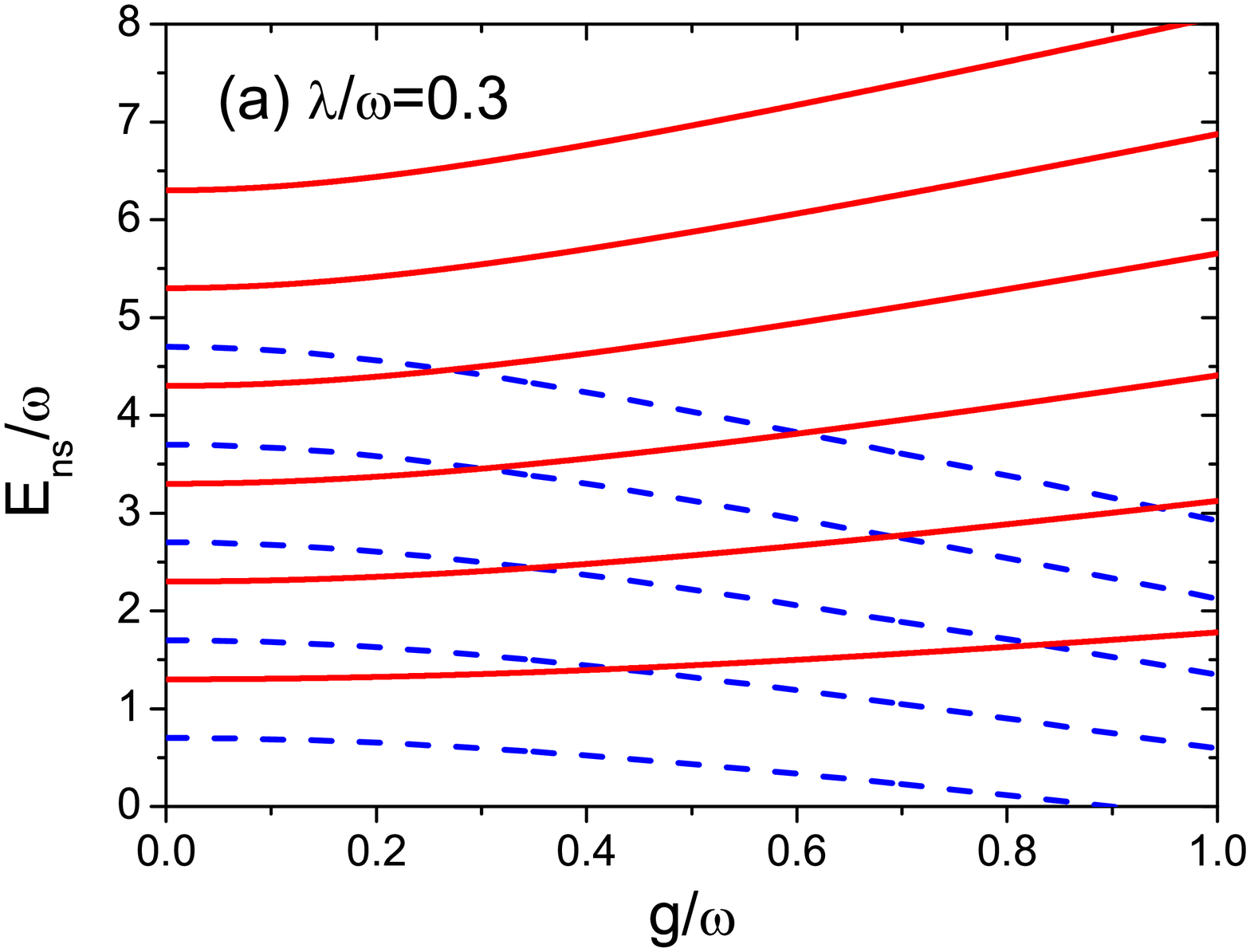}}
\rotatebox[origin=c]{0}{\includegraphics[angle=0,
          height=1.24in]{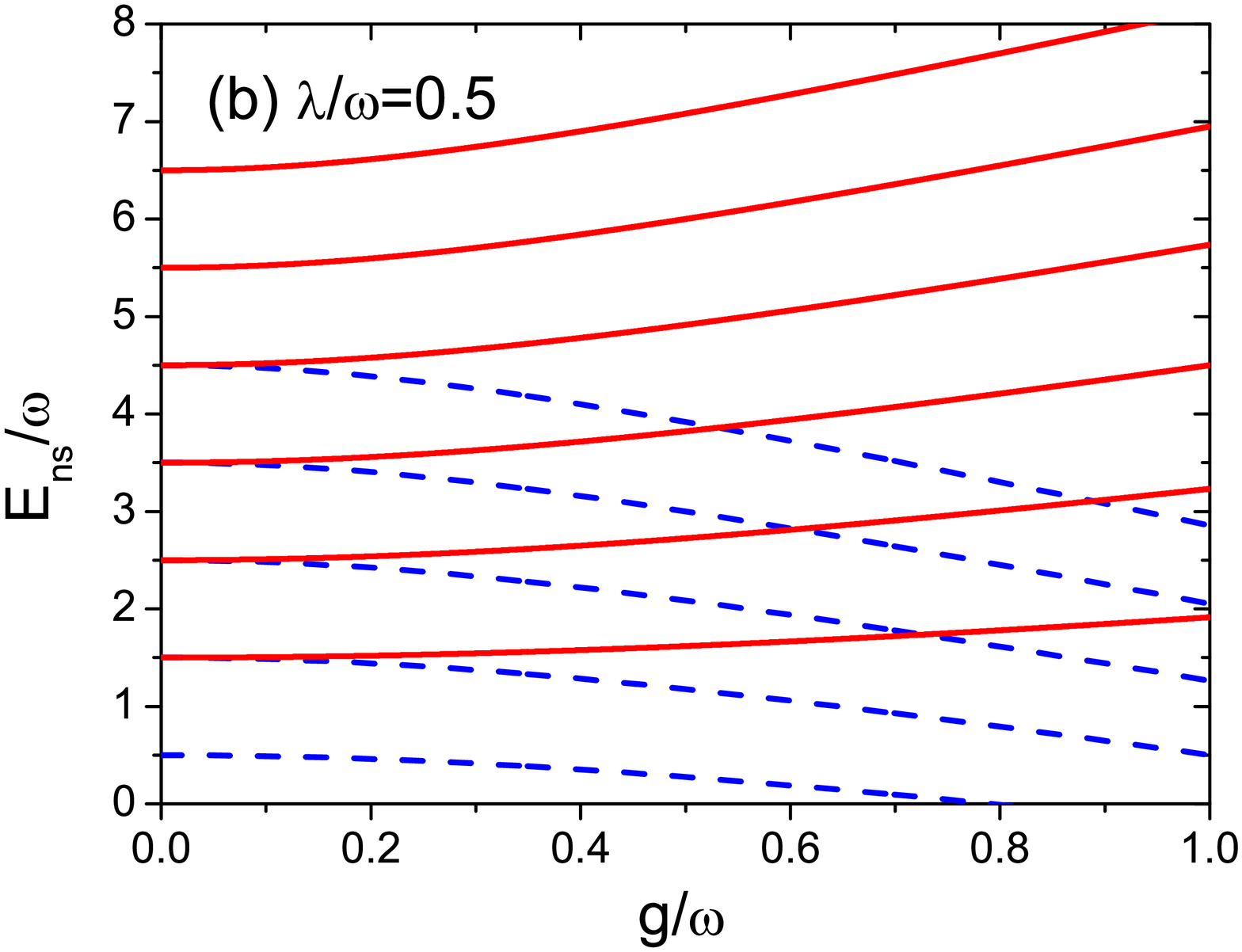}}
\caption {(Color online) The low-lying energy levels of the energy spectrum (13) in unit of $\omega$ as a function of the coupling parameter $g$ at different $\lambda$ under $\epsilon=0$. The solid lines denote $n=0, 1,\cdots, 5$ and $s=1$ while the dash lines mean $n=1, 2, \cdots, 5$ and $s=-1$.}
\end{figure}

When $\epsilon=0$, then $\Delta_{ns}=0$ according to Eq. (12). Therefore, the eigenvalue (9) has a simple formula
$$E_{ns}=(n+\frac{1}{2})\omega+s\sqrt{(\frac{\omega}{2}+\lambda)^2+(n+1)g^2}
\eqno{(13)}$$
in the absence of the driving term $\epsilon$. Obviously, the eigenvalue (13) recovers the exact solution of
the Hamiltonian (1) with $g=0$ and $\epsilon=0$. Based on the expression (13), we plot
the low-lying energy levels as a function of $g$ at different $\lambda$ in Fig. 1.
It is shown that there are level crossings between the neighboring eigenstates.
With increasing $\lambda$, the energy levels with $s=1 (-1)$ become higher (lower),
and these crossing points move toward the origin.

When $\epsilon\not=0$, $\Delta_{ns}$ in Eq. (12) with $\sigma=1$ has
an $\omega$-dependent solution. The corresponding eigenvalues $E_{ns}$ ($n=0, 1, 2,\cdots, \infty, s=\pm 1$) form
the sub-energy spectrum I. Fig. 2 depicts the low-lying energy levels  and the corresponding parameter
$\Delta_{ns}$ of the sub-energy spectrum I as a function of $g$ at different $\lambda$ under $\epsilon=0.4\omega$.
We note that when $g=0$, the sub-energy spectrum I becomes the exact eigenvalues for the interactionless case, i.e.
$E_{ns}=\omega n+s\sqrt{\lambda^2+\epsilon^2}$ with $\Delta_{ns}=(\lambda-\sqrt{\lambda^2+\epsilon^2})/\epsilon$.
Another $\omega$-dependent solution $\Delta_{ns}$ in Eq. (12) with $\sigma=-1$ is nothing but the sub-energy
spectrum II, which analytical expression is presented in the next subsection {\bf B}.

\begin{figure}
\rotatebox[origin=c]{0}{\includegraphics[angle=0,
          height=1.24in]{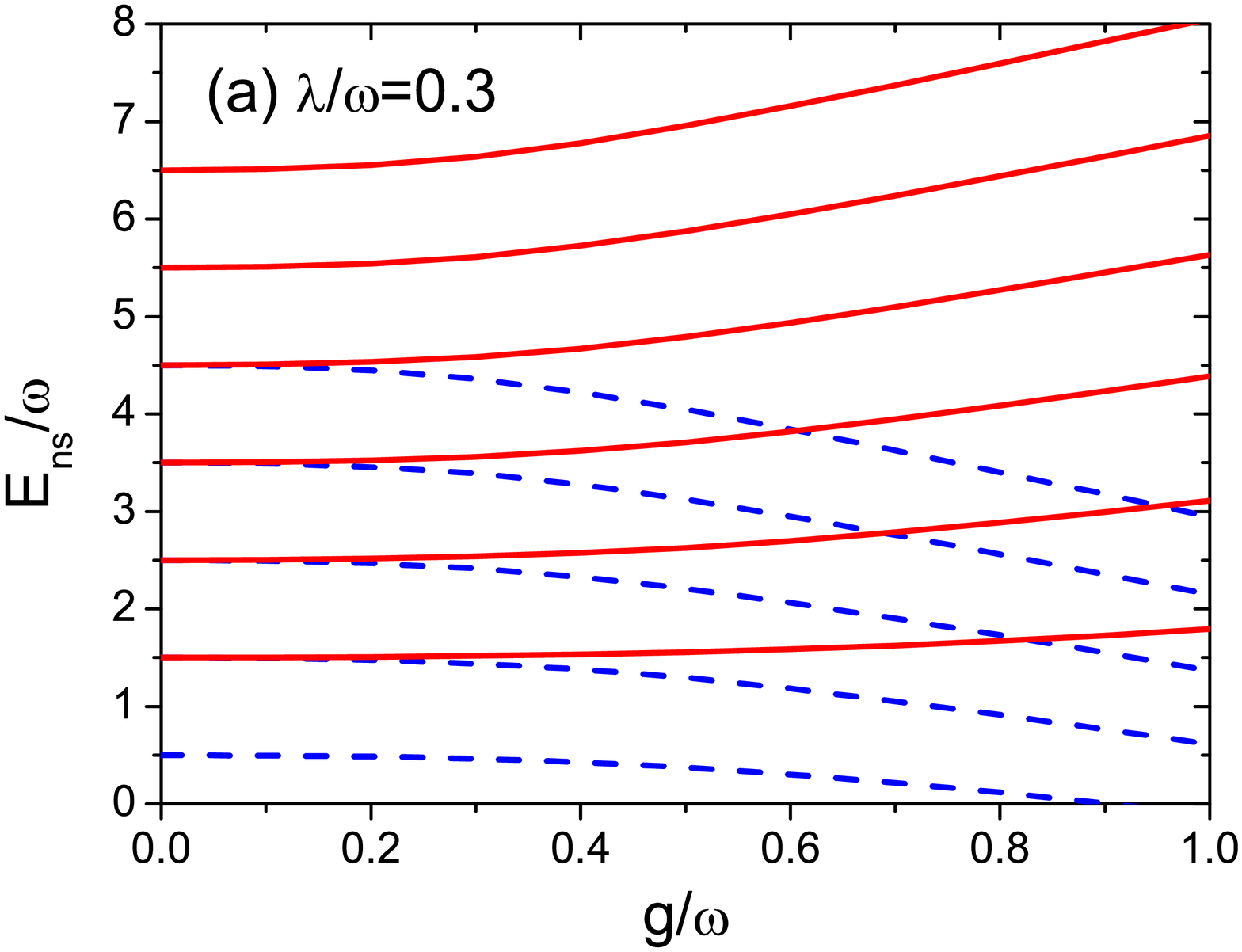}}
\rotatebox[origin=c]{0}{\includegraphics[angle=0,
          height=1.24in]{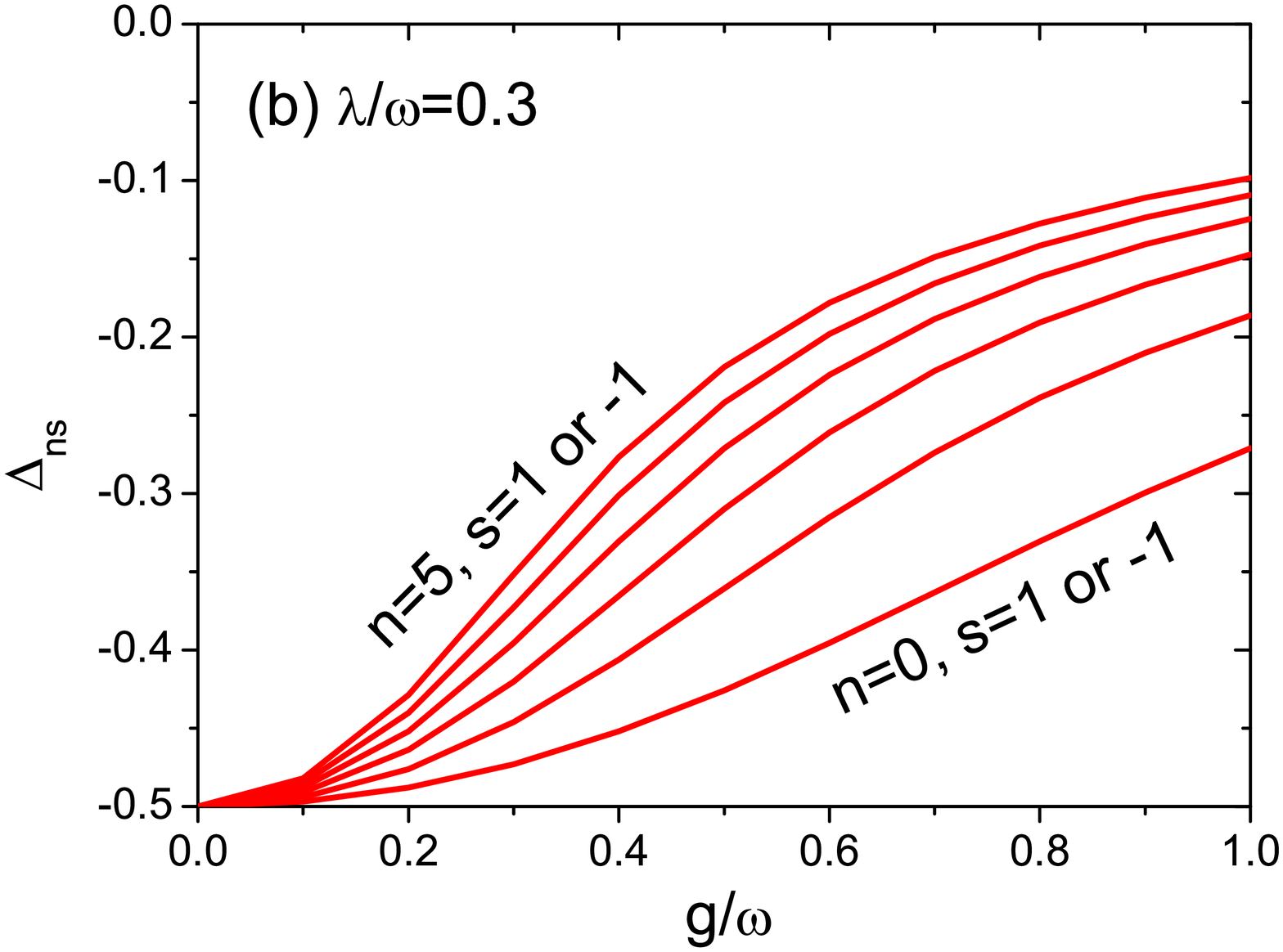}}
\rotatebox[origin=c]{0}{\includegraphics[angle=0,
          height=1.24in]{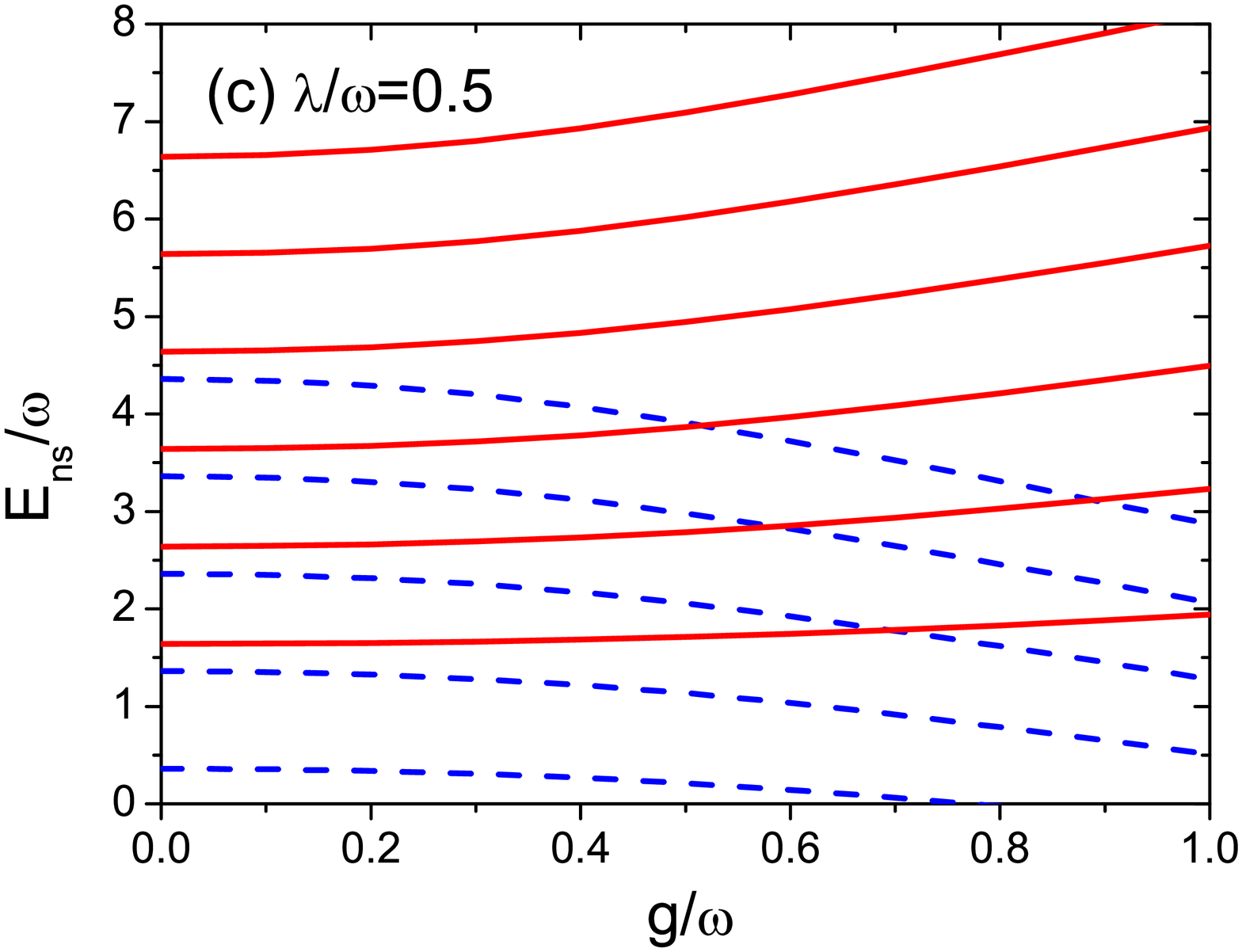}}
\rotatebox[origin=c]{0}{\includegraphics[angle=0,
          height=1.24in]{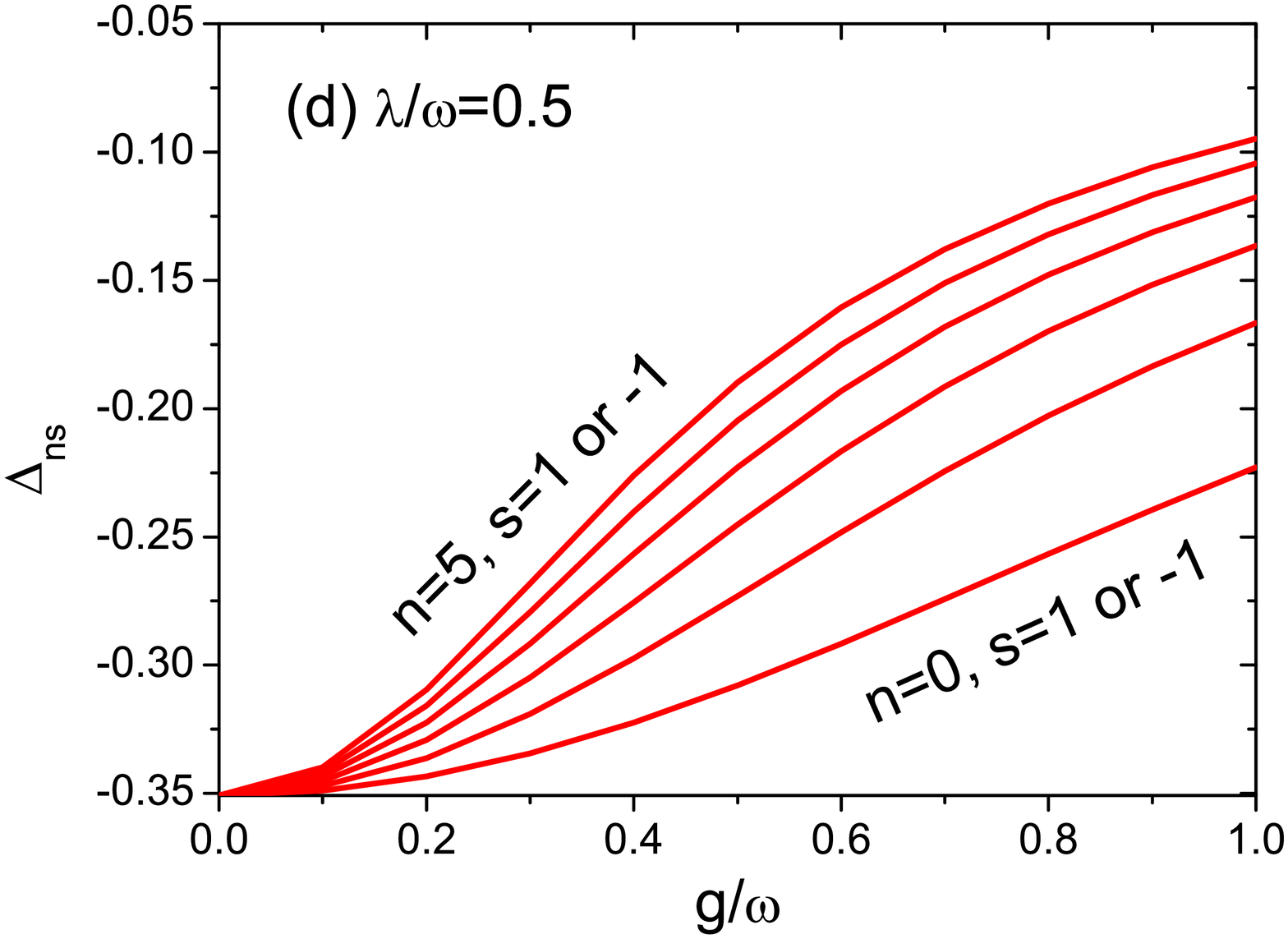}}
\caption {(Color online) The low-lying energy levels of the sub-energy spectrum I in unit of $\omega$ as a function of the coupling parameter $g$ at different $\lambda$ under $\epsilon=0.4\omega$, shown in (a) and (c). The solid lines denote $n=0, 1,\cdots, 5$ and $s=1$ while the dash lines mean $n=1, 2, \cdots, 5$ and $s=-1$. The corresponding $\Delta_{ns}$ are displayed in (b) and (d), respectively.}
\end{figure}

For the eigenstate associated with the sub-energy spectrum I, from Eq. (6), we have
$$\alpha^{ns}_{n+1}=\frac{(1+\Delta_{ns}^2)(E_{ns}-n\omega)+(1-\Delta_{ns}^2)\lambda-2\Delta_{ns}\epsilon}{g\sqrt{n+1}(1-\Delta_{ns}^2)}
\beta^{ns}_n,\eqno{(14)}$$
where $\beta^{ns}_n$ is an arbitrary constant and can be set to 1,
and the coefficients $\alpha^{ns}_{m}$ and $\beta^{ns}_{m}$  are uniquely determined by the recursion relations
$$\sqrt{m}
\left(
\begin{array}{c}
\alpha^{ns}_{m-1}\\
\beta^{ns}_{m-1}
\end{array}\right)
=-{\cal M}_{ns}^{-1}{\cal N}^{ns}_m\left (
\begin{array}{c}
\alpha^{ns}_{m}\\
\beta^{ns}_{m}
\end{array}\right )
-\sqrt{m+1}
\left(
\begin{array}{c}
\alpha^{ns}_{m+1}\\
\beta^{ns}_{m+1}
\end{array}\right)\eqno{(15)}$$
for $m=0, 1, 2, \cdots, n$, and
$$\sqrt{m+1}
\left(
\begin{array}{c}
\alpha^{ns}_{m+1}\\
\beta^{ns}_{m+1}
\end{array}\right)
=-{\cal M}_{ns}^{-1}{\cal N}^{ns}_m\left (
\begin{array}{c}
\alpha^{ns}_{m}\\
\beta^{ns}_{m}
\end{array}\right )
-\sqrt{m}
\left(
\begin{array}{c}
\alpha^{ns}_{m-1}\\
\beta^{ns}_{m-1}
\end{array}\right)\eqno{(16)}$$
for $m=n+1, n+2, \cdots, +\infty$.
Here we have defined
$$\begin{array}{lll}
{\cal M}_{ns}&=&\frac{g}{1+\Delta_{ns}^2}
[(1-\Delta_{ns}^2)\sigma_x-2\Delta_{ns}\sigma_z],\\
{\cal N}^{ns}_m&=&
(\omega m-E_{ns})I+(\frac{2\Delta_{ns}}{1+\Delta_{ns}^2}\lambda+\frac{1-\Delta_{ns}^2}{1+\Delta_{ns}^2}\epsilon)\sigma_x\\
&&+(\frac{1-\Delta_{ns}^2}{1+\Delta_{ns}^2}\lambda-\frac{2\Delta_{ns}}{1+\Delta_{ns}^2}\epsilon)\sigma_z,
\end{array}\eqno{(17)}$$
where $I$ is the $2\times 2$ unit matrix. From the recursion equation (15), we can see that
$\alpha^{ns}_{m-1}$ and $\beta^{ns}_{m-1}$ ($m=1, 2, \cdots, n$) are linear functions of
$\alpha^{ns}_{n}$ and $\beta^{ns}_{n+1}$, which are obtained by solving Eq. (15) with $m=0$.

\subsection{B. Sub-energy spectrum II}

Now we take another choice

$$\begin{array}{rrr}
(\omega n+\frac{1-\Delta^2_{ns}}{1+\Delta^2_{ns}}\lambda-\frac{2\Delta_{ns}}{1+\Delta^2_{ns}}\epsilon-E_{ns})
 \alpha^{ns}_{n}&&\\
+\frac{1-\Delta^2_{ns}}{1+\Delta^2_{ns}}g\sqrt{n+1}\beta^{ns}_{n+1}&=&0,\\
\end{array}
\eqno{(18)}$$
$$\begin{array}{rrr}
[\omega (n+1)-\frac{1-\Delta^2_{ns}}{1+\Delta^2_{ns}}\lambda+\frac{2\Delta_{ns}}{1+\Delta^2_{ns}}\epsilon-E_{ns}]
\beta^{ns}_{n+1}&&\\
+\frac{1-\Delta^2_{ns}}{1+\Delta^2_{ns}}g\sqrt{n+1}\alpha^{ns}_{n}&=&0\\
\end{array}
\eqno{(19)}$$
from the eigen-equations (3) and (4). Eqs. (18) and (19) originate in the vanishing of the two terms about $\alpha^{ns}_{n}$
and $\beta^{ns}_{n+1}$ in Eq. (3) with $m=n$ and Eq. (4) with $m=n+1$, respectively.
The corresponding eigenstate is uniquely determined by the constraint
$$[2\lambda\Delta_{ns}+\epsilon(1-\Delta^2_{ns})]\alpha^{ns}_n
+2g\Delta_{ns}\sqrt{n+1}\beta^{ns}_{n+1}=0,
\eqno{(20)}$$
or
$$[2\lambda\Delta_{ns}+\epsilon(1-\Delta^2_{ns})]\beta^{ns}_{n+1}
-2g\Delta_{ns}\sqrt{n+1}\alpha^{ns}_n=0,
\eqno{(21)}$$
Solving Eqs. (18) and (19), we obtain
$$\begin{array}{l}
E_{ns}=(n+\frac{1}{2})\omega+s\Theta_{ns},\\
\Theta_{ns}=\sqrt{(\frac{\omega}{2}-\frac{\lambda(1-\Delta^2_{ns})-2\epsilon\Delta_{ns}}{1+\Delta^2_{ns}}
)^2+(n+1)g^2(\frac{1-\Delta^2_{ns}}{1+\Delta^2_{ns}})^2}.
\end{array}
\eqno{(22)}$$
Here $\Delta_{ns}$ satisfies
$$\epsilon(1+\Delta_{ns}^2)+2\Delta_{ns}(E_{ns}-\omega n)=0,
\eqno{(23)}$$
or
$$\epsilon(1+\Delta_{ns}^2)-2\Delta_{ns}[E_{ns}-\omega (n+1)]=0,
\eqno{(24)}$$
which is derived from Eqs. (18) and (20) or Eqs. (19) and (21), respectively.
Similar to Eqs. (10) and (11) in the previous subsection {\bf A}, Eq. (23) with $s=-1 (1)$
is also consistent with Eq. (24) with $s=1 (-1)$. This leads to the parameter equation
$$\epsilon(1+\Delta_{ns}^2)+\Delta_{ns}(2\tau \Theta_{ns}+\omega)=0,
\eqno{(25)}$$
where $\tau=\pm 1$.

If $\epsilon=0$, then $\Delta_{ns}=0$ from Eq. (25).
So the eigenvalue (22) also has an explicit expression
$$E_{ns}=(n+\frac{1}{2})\omega+s\sqrt{(\frac{\omega}{2}-\lambda)^2+(n+1)g^2}.
\eqno{(26)}$$
We depict the low-lying energy levels as a function of $g$ at $\lambda=0.3\omega$, $0.5\omega$, and $\epsilon=0$ in Fig. 3.

\begin{figure}
\rotatebox[origin=c]{0}{\includegraphics[angle=0,
          height=1.24in]{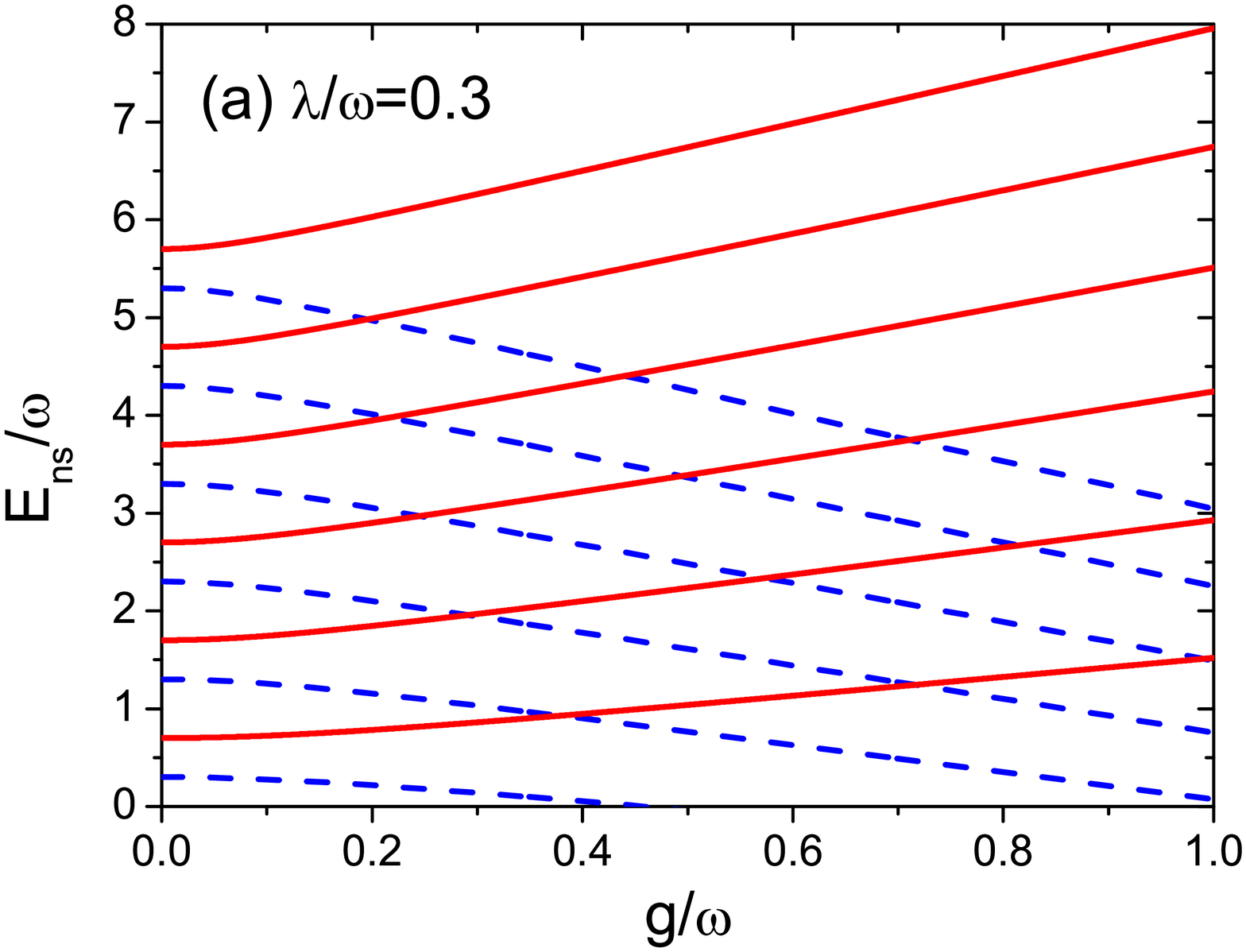}}
\rotatebox[origin=c]{0}{\includegraphics[angle=0,
          height=1.24in]{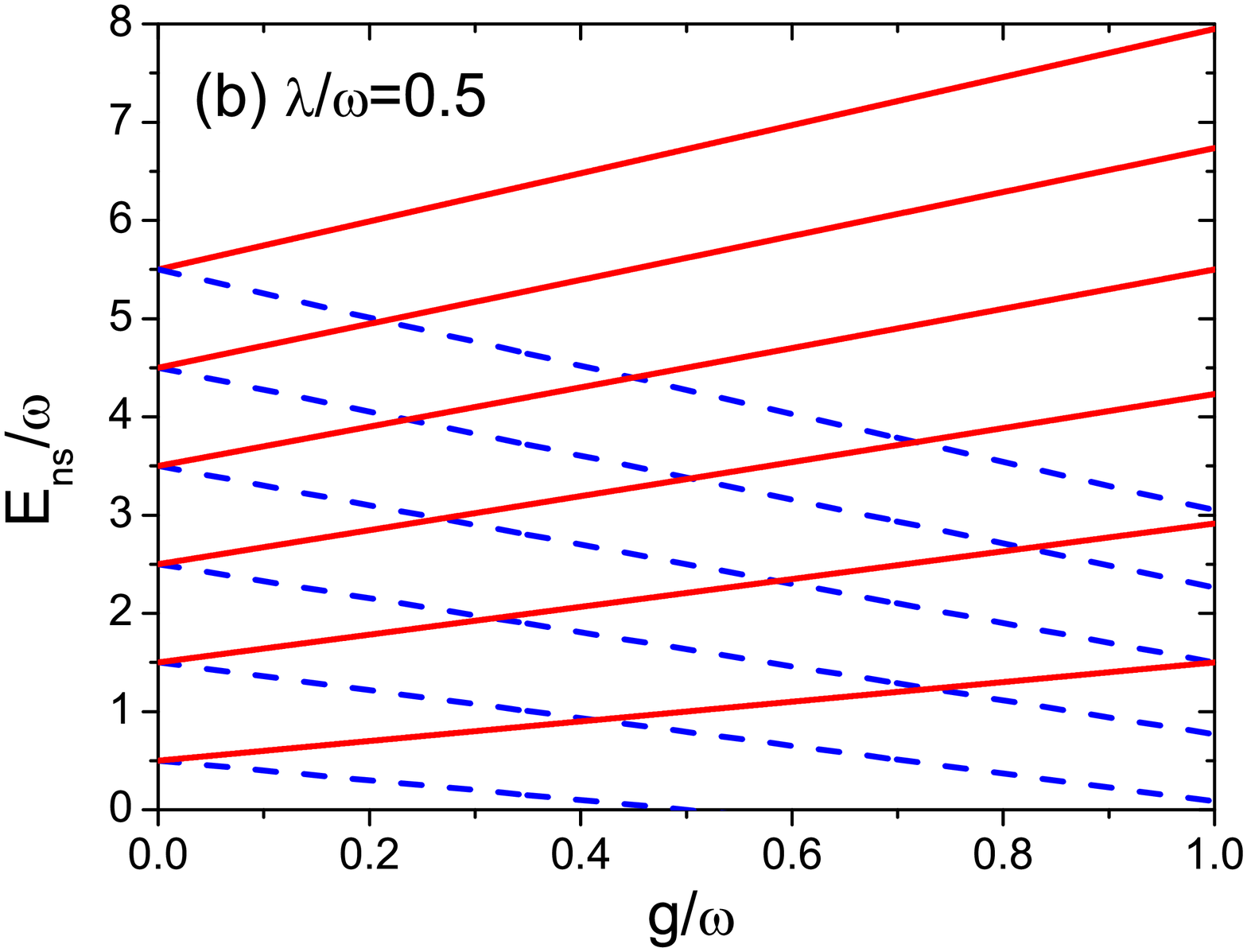}}
\caption {(Color online) The low-lying energy levels of the energy spectrum (26) in unit of $\omega$ as a function of the coupling parameter $g$ at different $\lambda$ under $\epsilon=0$. The solid lines denote $n=0, 1,\cdots, 5$ and $s=1$ while the dash lines mean $n=0, 1, \cdots, 5$ and $s=-1$.}
\end{figure}

When $\epsilon\not=0$, $\Delta_{ns}$ in Eq. (25) with $\tau=1$ also has
an $\omega$-dependent solution. The corresponding eigenvalues constitute
the sub-energy spectrum II. Fig. 4 exhibits the low-lying energy levels of the sub-energy spectrum II and the corresponding parameter
$\Delta_{ns}$ as a function of $g$ at $\lambda=0.3\omega$, $0.5\omega$, and $\epsilon=0.4\omega$.
We note that after taking the transformation $\Delta_{ns}\rightarrow -1/\Delta_{ns}$,
the eigenvalue (22) with $\tau=1(-1)$ becomes  the eigenvalue (10) with $\sigma=-1(1)$.
Therefore, both the sub-energy spectrum I and II are double degenerate.

\begin{figure}
\rotatebox[origin=c]{0}{\includegraphics[angle=0,
          height=1.24in]{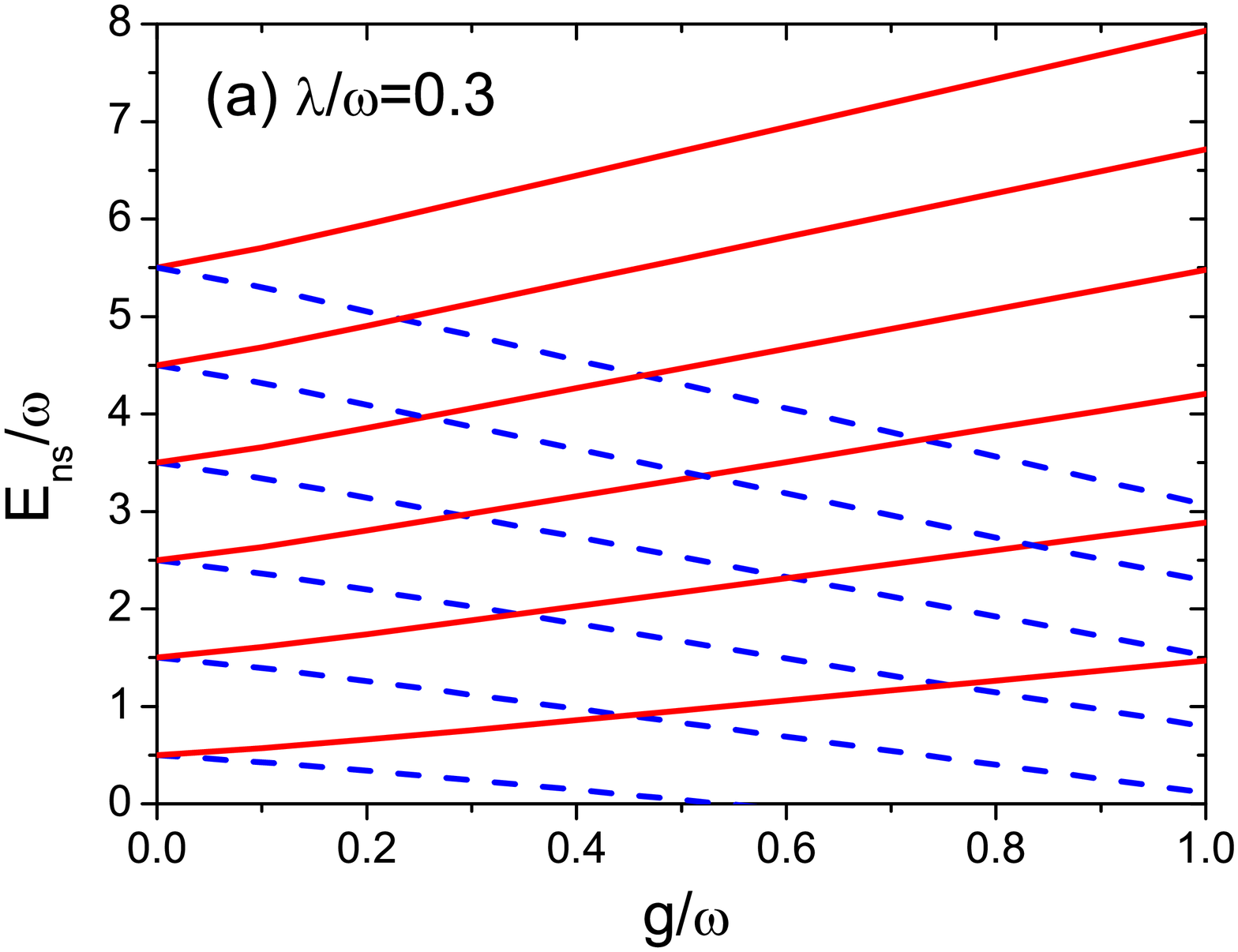}}
\rotatebox[origin=c]{0}{\includegraphics[angle=0,
          height=1.24in]{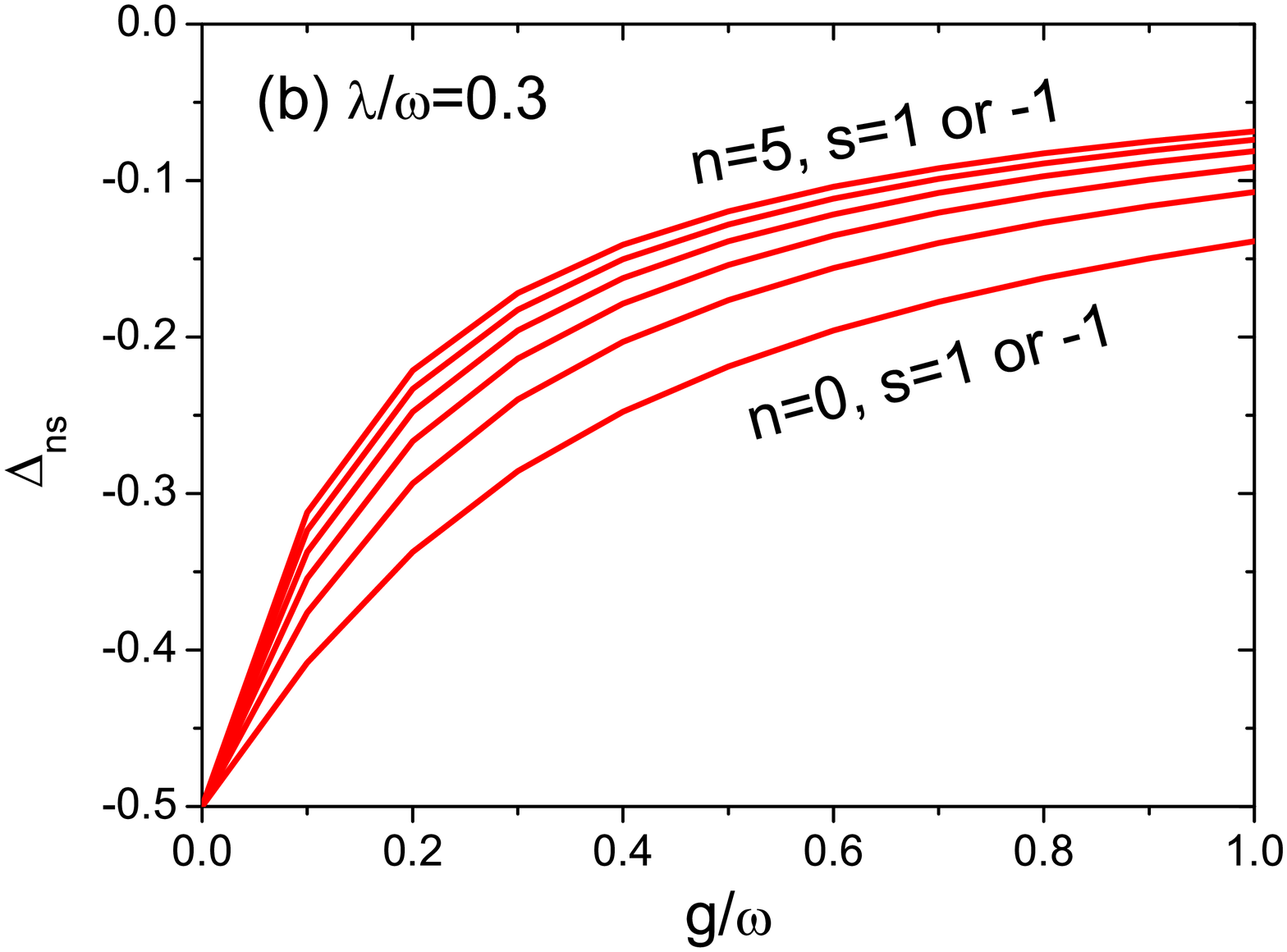}}
\rotatebox[origin=c]{0}{\includegraphics[angle=0,
          height=1.24in]{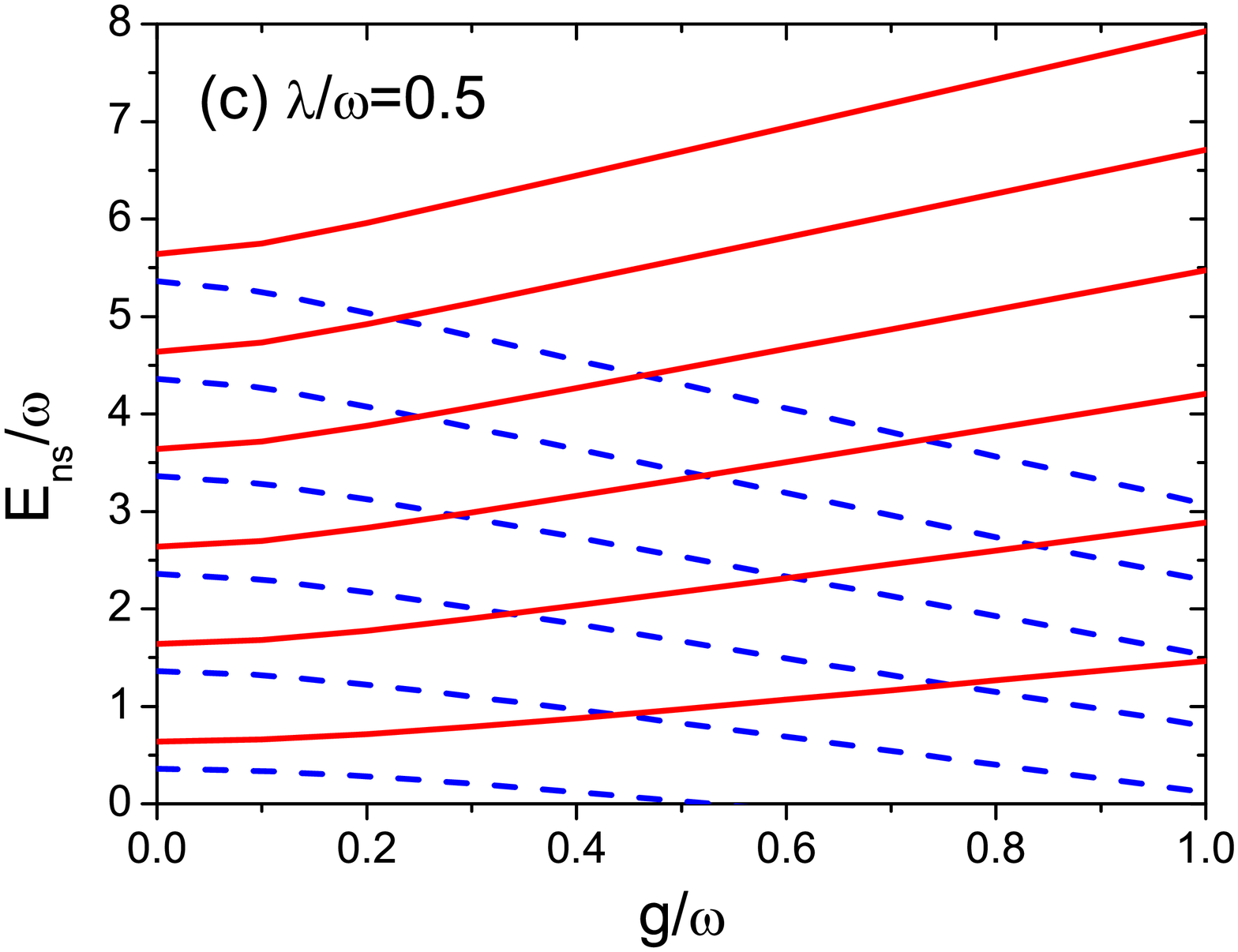}}
\rotatebox[origin=c]{0}{\includegraphics[angle=0,
          height=1.24in]{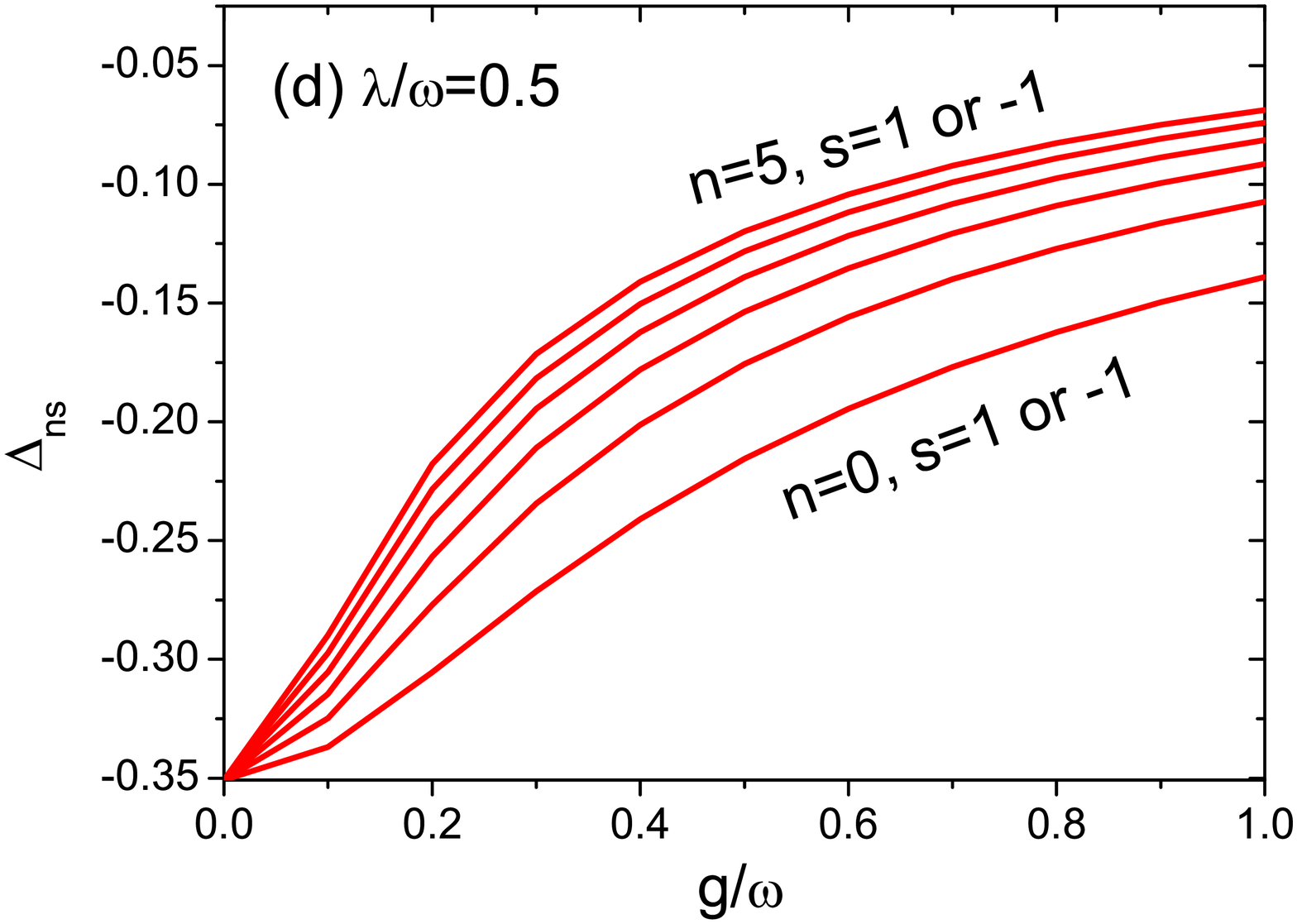}}
\caption {(Color online) The low-lying energy levels of the sub-energy spectrum II in unit of $\omega$ as a function of the coupling parameter $g$ at different $\lambda$ under $\epsilon=0.4\omega$, shown in (a) and (c). The solid lines denote $n=0, 1,\cdots, 5$ and $s=1$ while the dash lines mean $n=0, 1, \cdots, 5$ and $s=-1$. The corresponding $\Delta_{ns}$ are displayed in (b) and (d), respectively.}
\end{figure}

For the nth eigenstate with $s$ in the sub-energy spectrum II, we have
$$\beta^{ns}_{n+1}=\frac{(1+\Delta_{ns}^2)(E_{ns}-n\omega)-(1-\Delta_{ns}^2)\lambda+2\Delta_{ns}\epsilon}{g\sqrt{n+1}(1-\Delta_{ns}^2)}
\alpha^{ns}_n,\eqno{(27)}$$
where $\alpha^{ns}_n$ is an arbitrary constant and is set to 1.
The other coefficients $\alpha^{ns}_i$ and $\beta^{ns}_i$ also obey
the same recursion relations (15) and (16) in the sub-energy spectrum I.

\section{III. The Bargmann space}

In this section, we reinvestigate the eigenvalue problem for the Hamiltonian (1) in the Bargmann space [16],
where the bosonic creation and anihilation operators in terms of a complex variable $z$ can be transformed as
$a^\dagger\rightarrow z$ and $a\rightarrow d/dz$, respectively.
Then the Hamiltonian (1) becomes
$$ H=\left (
\begin{array}{cc}
\omega z\frac{d}{dz}+\lambda & g(z+\frac{d}{dz})+\epsilon\\
g(z+\frac{d}{dz})+\epsilon & \omega z\frac{d}{dz}-\lambda \\
\end{array}\right ).
\eqno{(28)} $$
In this representation, the state $\Psi(z)$ can be normalized according to
$$<\Psi|\Psi>=\frac{1}{\pi}\int dzd\overline{z}e^{-z\overline{z}}\Psi^+(z)\Psi(z)\equiv 1.\eqno{(29)}$$

We assume that the two-component eigenstate of the Hamiltonian (28) for the $n$th energy level
with quantum number $s$ possesses the general form
$$\Psi_{ns}=\sum_{i=0}^{+\infty}\frac{1}{\sqrt{1+\Delta^2_{ns}}}
\left(
\begin{array}{cc}
1&\Delta_{ns}\\
-\Delta_{ns}&1
\end{array}\right)
\left (
\begin{array}{c}
A^{ns}_{i}z^i\\
B^{ns}_{i}z^i
\end{array}\right ),\eqno{(30)}$$
where $s=\pm 1$, $\Delta_{ns}$ is a real parameter in the unitary matrix to be determined below
by requiring the coefficients $A^{ns}_{i}$ and $B^{ns}_{i}$ to be nonzero. When $i \rightarrow + \infty$,
$A^{ns}_{i}\rightarrow 0$ and $B^{ns}_{i}\rightarrow 0$, so that $\Psi_{ns}$ is finite at any $z$ in the Bargmann space.
Substituting the eigenfunction (30) into the eigen-equation $H\Psi_{ns}=E_{ns}\Psi_{ns}$ and
requiring the coefficients of $z^i$ to be zero, we obtain
the infinite system of homogeneous linear equations with the variables $A^{ns}_{i}$ and $B^{ns}_{i}$
$$\begin{array}{rrr}
\frac{2g\Delta_{ns}}{1+\Delta_{ns}^2}A^{ns}_{i-1}+[E_{ns}-i\omega-\frac{\lambda(1-\Delta^2_{ns})-2\epsilon\Delta_{ns}}{1+\Delta_{ns}^2}]A^{ns}_i& &\\
+\frac{2g\Delta_{ns}(i+1)}{1+\Delta_{ns}^2}A^{ns}_{i+1}-\frac{g(1-\Delta^2_{ns})}{1+\Delta_{ns}^2}B^{ns}_{i-1}& &\\
-\frac{2\lambda\Delta_{ns}+\epsilon(1-\Delta^2_{ns})}{1+\Delta_{ns}^2}B^{ns}_i-\frac{g(1-\Delta^2_{ns})(i+1)}{1+\Delta_{ns}^2}B^{ns}_{i+1}&=&0,\\
\end{array}\eqno{(31)}$$
$$\begin{array}{rrr}
\frac{2g\Delta_{ns}}{1+\Delta_{ns}^2}B^{ns}_{i-1}-[E_{ns}-i\omega+\frac{\lambda(1-\Delta^2_{ns})-2\epsilon\Delta_{ns}}{1+\Delta_{ns}^2}]B^{ns}_i& &\\
+\frac{2g\Delta_{ns}(i+1)}{1+\Delta_{ns}^2}B^{ns}_{i+1}+\frac{g(1-\Delta^2_{ns})}{1+\Delta_{ns}^2}A^{ns}_{i-1}& &\\
+\frac{2\lambda\Delta_{ns}+\epsilon(1-\Delta^2_{ns})}{1+\Delta_{ns}^2}A^{ns}_i+\frac{g(1-\Delta^2_{ns})(i+1)}{1+\Delta_{ns}^2}A^{ns}_{i+1}&=&0,\\
\end{array}\eqno{(32)}$$
where $i=0, 1, 2, \cdots, \infty, A^{ns}_{m}=B^{ns}_{m}\equiv0$ for $m<0$.
Eqs. (31) and (32) can be also solved exactly by employing the same procedure in the occupation number representation in section II.

\subsection{A. Sub-energy spectrum I}

Following the trick presented in the occupation number representation,
we let
$$\begin{array}{rrr}
[\omega(n+1)+\frac{\lambda(1-\Delta^2_{ns})-2\epsilon\Delta_{ns}}{1+\Delta^2_{ns}}-E_{ns}]A^{ns}_{n+1}&&\\
+\frac{g(1-\Delta^2_{ns})}{1+\Delta^2_{ns}}B^{ns}_n&=&0,\\
\end{array}
\eqno{(33)}$$
$$\begin{array}{rrr}
[\omega n-\frac{\lambda(1-\Delta^2_{ns})-2\epsilon\Delta_{ns}}{1+\Delta^2_{ns}}-E_{ns}]B^{ns}_n&&\\
+\frac{g(1-\Delta^2_{ns})(n+1)}{1+\Delta^2_{ns}}A^{ns}_{n+1}&=&0,\\
\end{array}
\eqno{(34)}$$
which come from the vanishing of the two terms about $A^{ns}_{n+1}$ and $B^{ns}_n$
in Eq. (31) with $i=n+1$ and Eq.(32) with $i=n$, respectively.
Then the non-zero eigenfunction associated with the eigenvalue $E_{ns}$ is solely fixed by requiring
$$[2\lambda\Delta_{ns}+\epsilon(1-\Delta^2_{ns})]B^{ns}_n-2g\Delta_{ns}(n+1)A^{ns}_{n+1}=0,
\eqno{(35)}$$
or
$$2g\Delta_{ns}B^{ns}_n+[2\lambda\Delta_{ns}+\epsilon(1-\Delta^2_{ns})]A^{ns}_{n+1}=0.
\eqno{(36)}$$
Solving the homogenous linear equations (33) and (34) about $A^{ns}_{n+1}$
and $B^{ns}_n$, we have
$$\begin{array}{l}
E_{ns}=(n+\frac{1}{2})\omega+s \Xi_{ns}, \\
\Xi_{ns}=\sqrt{(\frac{\omega}{2}+\frac{\lambda(1-\Delta^2_{ns})-2\epsilon\Delta_{ns}}{1+\Delta^2_{ns}}
)^2+(n+1)g^2(\frac{1-\Delta^2_{ns}}{1+\Delta^2_{ns}})^2},
\end{array}
\eqno{(37)}$$
which is nothing but the eigenvalue (9) in the occupation number representation in section II.
Substituting  $A^{ns}_{n+1}$ in Eq. (34) into Eq. (35) or $B^{ns}_n$ in Eq. (33) into Eq. (36),
we obtain
$$\epsilon(1+\Delta_{ns}^2)-2\Delta_{ns}(E_{ns}-\omega n)=0,
\eqno{(38)}$$
or
$$\epsilon(1+\Delta_{ns}^2)+2\Delta_{ns}[E_{ns}-\omega (n+1)]=0.
\eqno{(39)}$$
Surprisingly, Eqs. (38) and (39) also coincide with Eqs. (10) and (11) in the occupation number representation,
respectively.

\begin{figure}
\rotatebox[origin=c]{0}{\includegraphics[angle=0,
          height=1.24in]{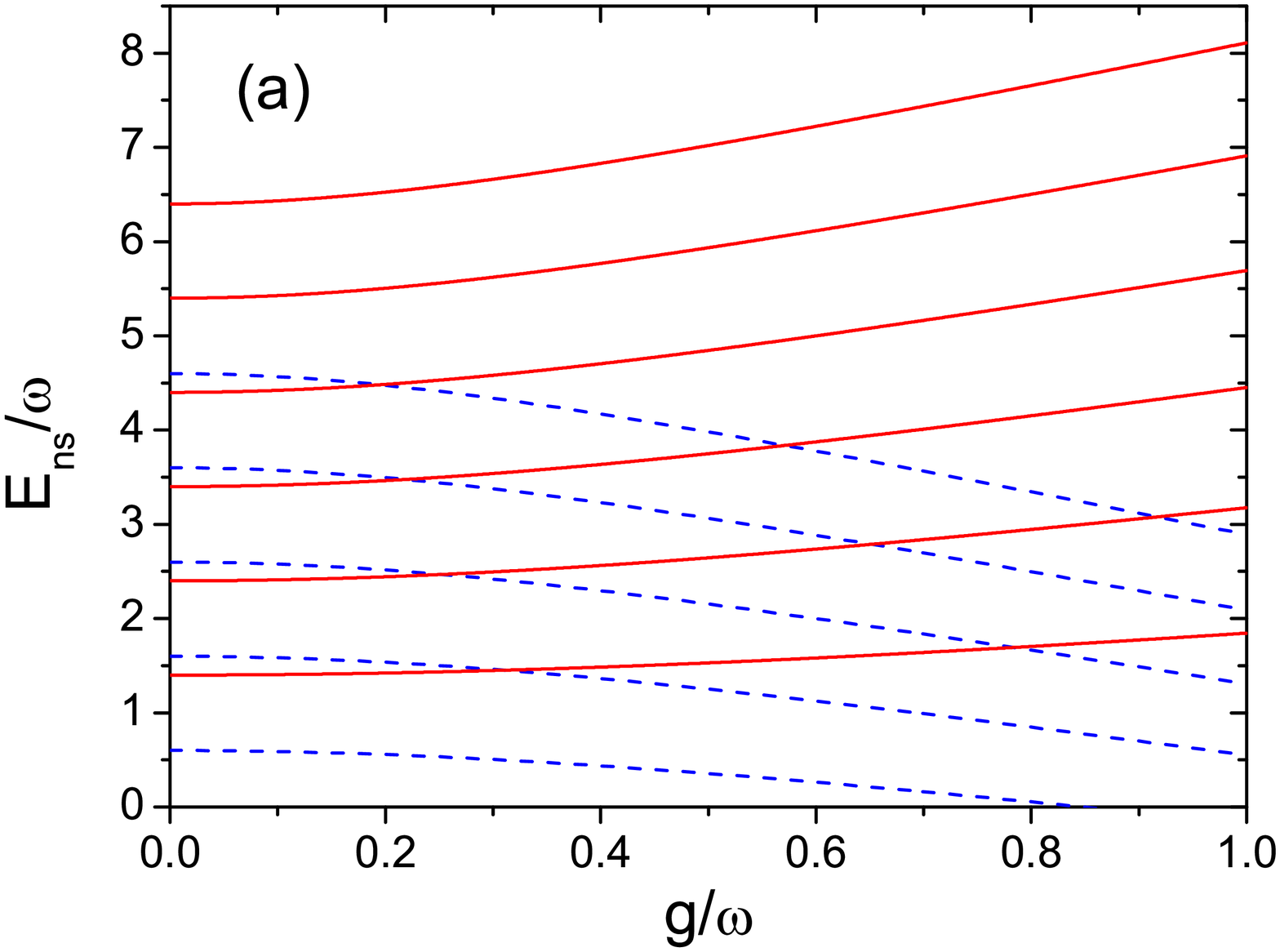}}
\rotatebox[origin=c]{0}{\includegraphics[angle=0,
          height=1.24in]{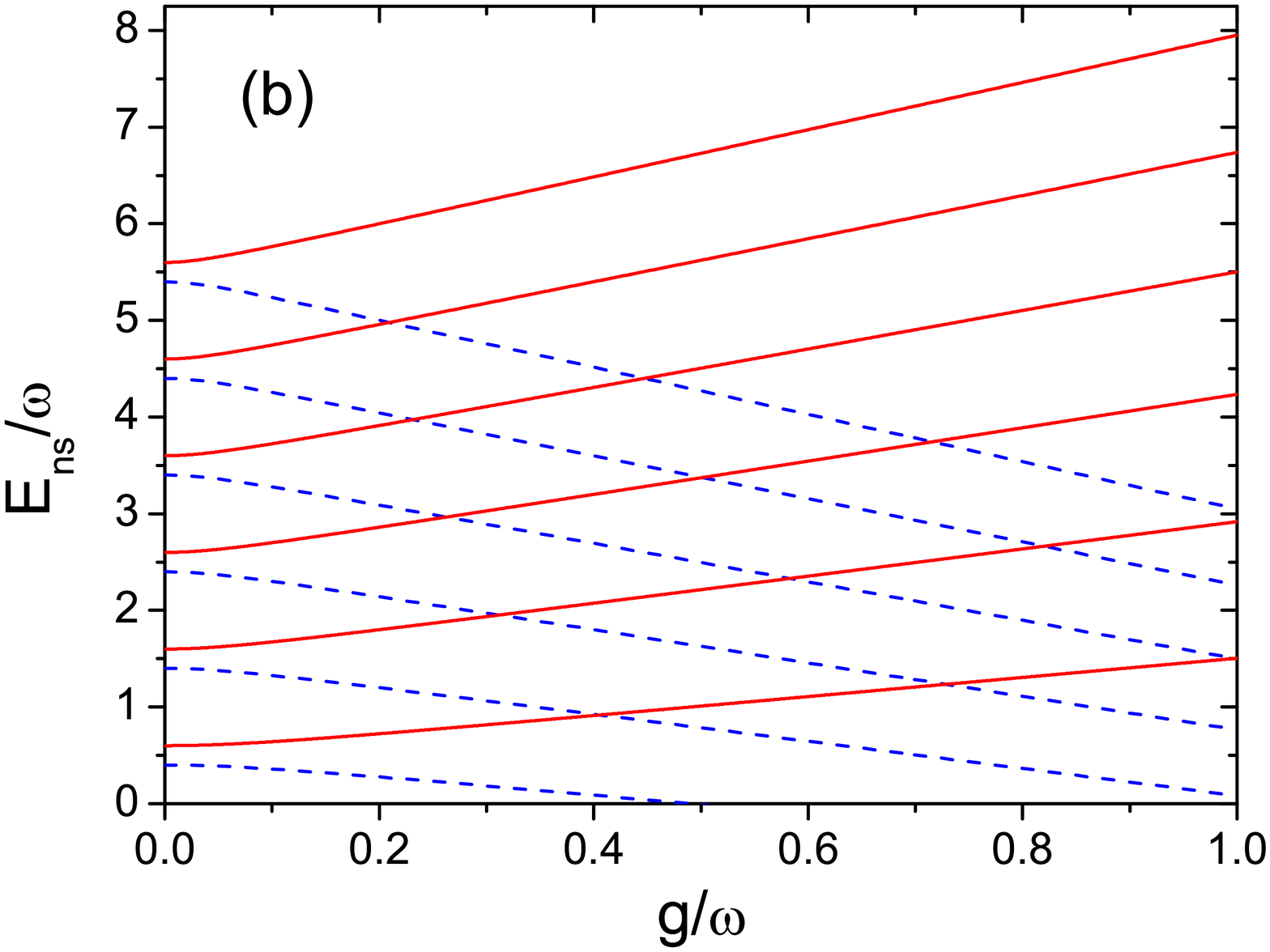}}
\caption {(Color online) The low-lying energy levels of the sub-energy spectrum I (a) and II (b) in unit of $\omega$ as a function of the coupling parameter $g$ when $\lambda=0.4\omega$ and $\epsilon=0$. The solid lines denote $n=0, 1,\cdots, 5$ and $s=1$ while the dash lines mean $n=1, 2, \cdots, 5$ and $s=-1$.}
\end{figure}

For the eigenstate for the sub-energy spectrum I, from Eq. (34), we have
$$A^{ns}_{n+1}=\frac{(1+\Delta_{ns}^2)(E_{ns}-n\omega)+\lambda(1-\Delta_{ns}^2)-2\epsilon\Delta_{ns}}{g(1-\Delta_{ns}^2)(n+1)}
B^{ns}_n,\eqno{(40)}$$
where $B^{ns}_n$ is a constant to be determined by the normalized condition (29).
The coefficients $\alpha^{ns}_{i}$ and $\beta^{ns}_{i}$, proportional to $B^{ns}_n$, are obtained
by the recursion relations
$$
\left(
\begin{array}{c}
A^{ns}_{i-1}\\
B^{ns}_{i-1}
\end{array}\right)
=-{\cal M}^{-1}_{ns}{\cal N}^{ns}_i\left (
\begin{array}{c}
A^{ns}_{i}\\
B^{ns}_{i}
\end{array}\right )
-(i+1)
\left(
\begin{array}{c}
A^{ns}_{i+1}\\
B^{ns}_{i+1}
\end{array}\right)\eqno{(41)}$$
for $i=0, 1, 2, \cdots, n$, and
$$
\left(
\begin{array}{c}
A^{ns}_{i+1}\\
B^{ns}_{i+1}
\end{array}\right)
=-\frac{{\cal M}^{-1}_{ns}{\cal N}^{ns}_i}{i+1}\left (
\begin{array}{c}
A^{ns}_{i}\\
B^{ns}_{i}
\end{array}\right )
-\frac{1}{i+1}
\left(
\begin{array}{c}
A^{ns}_{i-1}\\
B^{ns}_{i-1}
\end{array}\right)\eqno{(42)}$$
for $i=n+1, n+2, \cdots, +\infty$.

\subsection{B. Sub-energy spectrum II}

From the eigen-equations (31) and (32), we require
$$\begin{array}{rrr}
(\omega n+\frac{\lambda(1-\Delta^2_{ns})-2\epsilon\Delta_{ns}}{1+\Delta^2_{ns}}-E_{ns})A^{ns}_{n}&&\\
+\frac{g(1-\Delta^2_{ns})(n+1)}{1+\Delta^2_{ns}}B^{ns}_{n+1}&=&0,\\
\end{array}
\eqno{(43)}$$
$$\begin{array}{rrr}
[\omega (n+1)-\frac{\lambda(1-\Delta^2_{ns})-2\epsilon\Delta_{ns}}{1+\Delta^2_{ns}}\lambda-E_{ns}]B^{ns}_{n+1}&&\\
+\frac{g(1-\Delta^2_{ns})}{1+\Delta^2_{ns}}A^{ns}_{n}&=&0.\\
\end{array}
\eqno{(44)}$$
The equations above originate in the vanishing of the two terms about $A^{ns}_{n}$ and $B^{ns}_{n+1}$ in
Eq. (31) with $i=n$ and Eq. (32) with $i=n+1$, respectively.
The corresponding eigenfunction is uniquely determined by the condition
$$[2\lambda\Delta_{ns}+\epsilon(1-\Delta^2_{ns})]A^{ns}_n+2g\Delta_{ns}(n+1)B^{ns}_{n+1}=0,
\eqno{(45)}$$
or
$$-2g\Delta_{ns}A^{ns}_n+[2\lambda\Delta_{ns}+\epsilon(1-\Delta^2_{ns})]B^{ns}_{n+1}=0,
\eqno{(46)}$$
Solving Eqs. (43) and (44), we have
$$\begin{array}{l}
E_{ns}=(n+\frac{1}{2})\omega+s\Theta_{ns},\\
\Theta_{ns}=\sqrt{(\frac{\omega}{2}-\frac{\lambda(1-\Delta^2_{ns})-2\epsilon\Delta_{ns}}{1+\Delta^2_{ns}}
)^2+(n+1)g^2(\frac{1-\Delta^2_{ns}}{1+\Delta^2_{ns}})^2},
\end{array}
\eqno{(47)}$$
which is consistent with the eigenvalue (22) in the occupation number representation.
Here $\Delta_{ns}$ satisfies the nonlinear equation
$$\epsilon(1+\Delta_{ns}^2)+2\Delta_{ns}(E_{ns}-\omega n)=0,
\eqno{(48)}$$
or
$$\epsilon(1+\Delta_{ns}^2)-2\Delta_{ns}[E_{ns}-\omega (n+1)]=0,
\eqno{(49)}$$
which is derived from Eqs. (43) and (45) or Eqs. (44) and (46), respectively.
Obviously, Eqs. (48) and (49) are also identical to Eqs.(23) and (24)
in the occupation number representation, respectively.

\begin{figure}
\rotatebox[origin=c]{0}{\includegraphics[angle=0,
          height=1.24in]{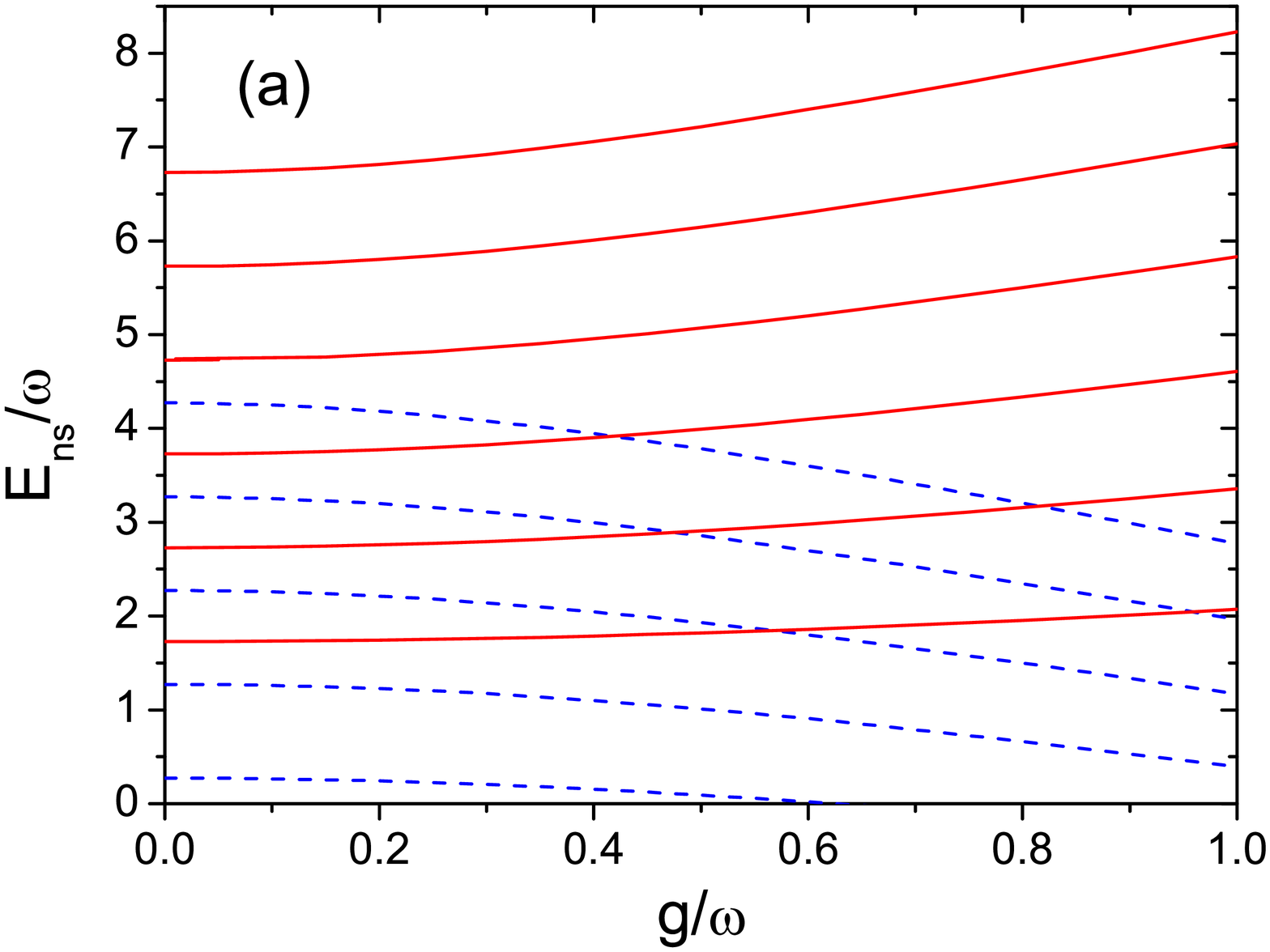}}
\rotatebox[origin=c]{0}{\includegraphics[angle=0,
          height=1.24in]{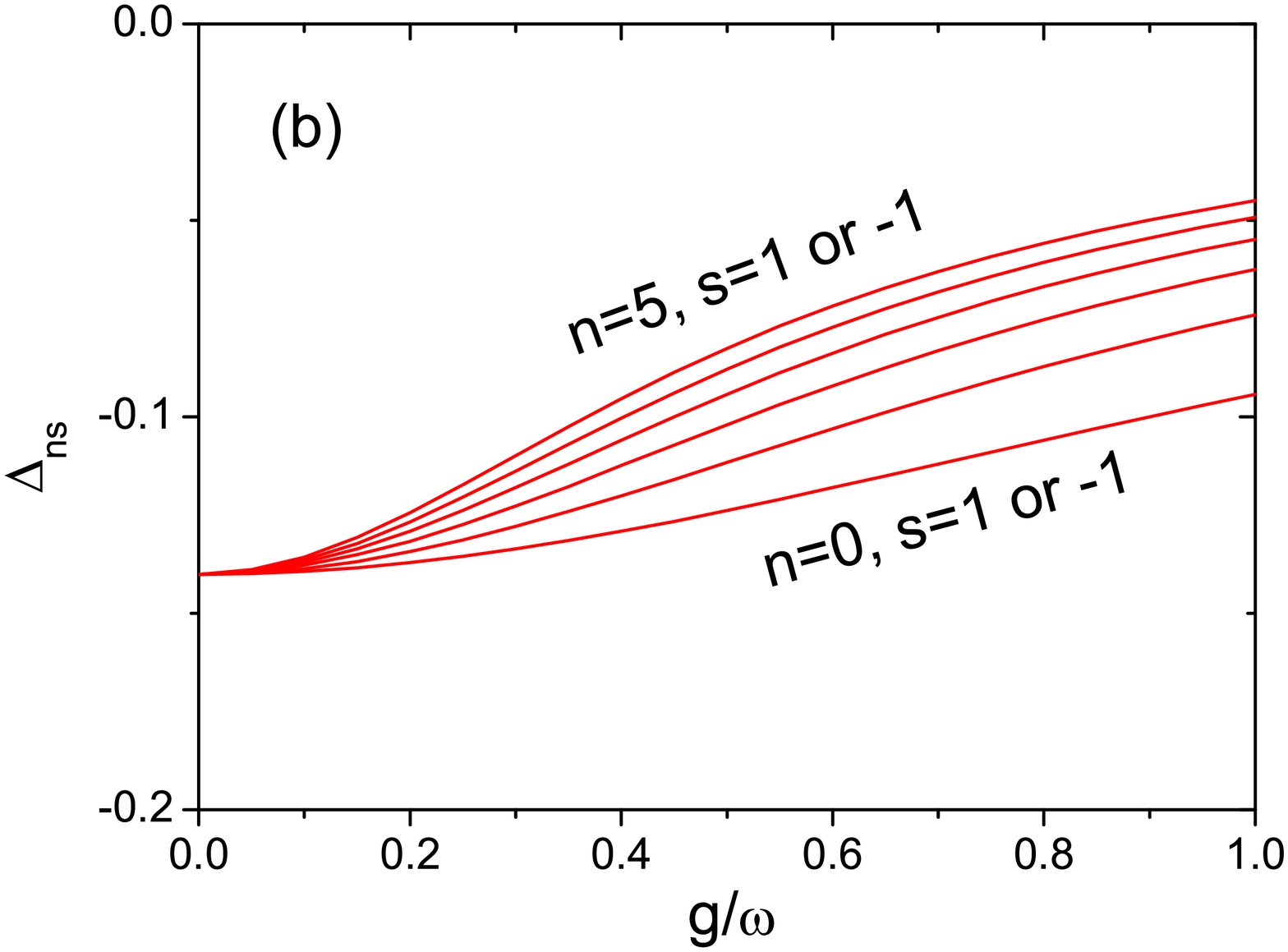}}
\rotatebox[origin=c]{0}{\includegraphics[angle=0,
          height=1.24in]{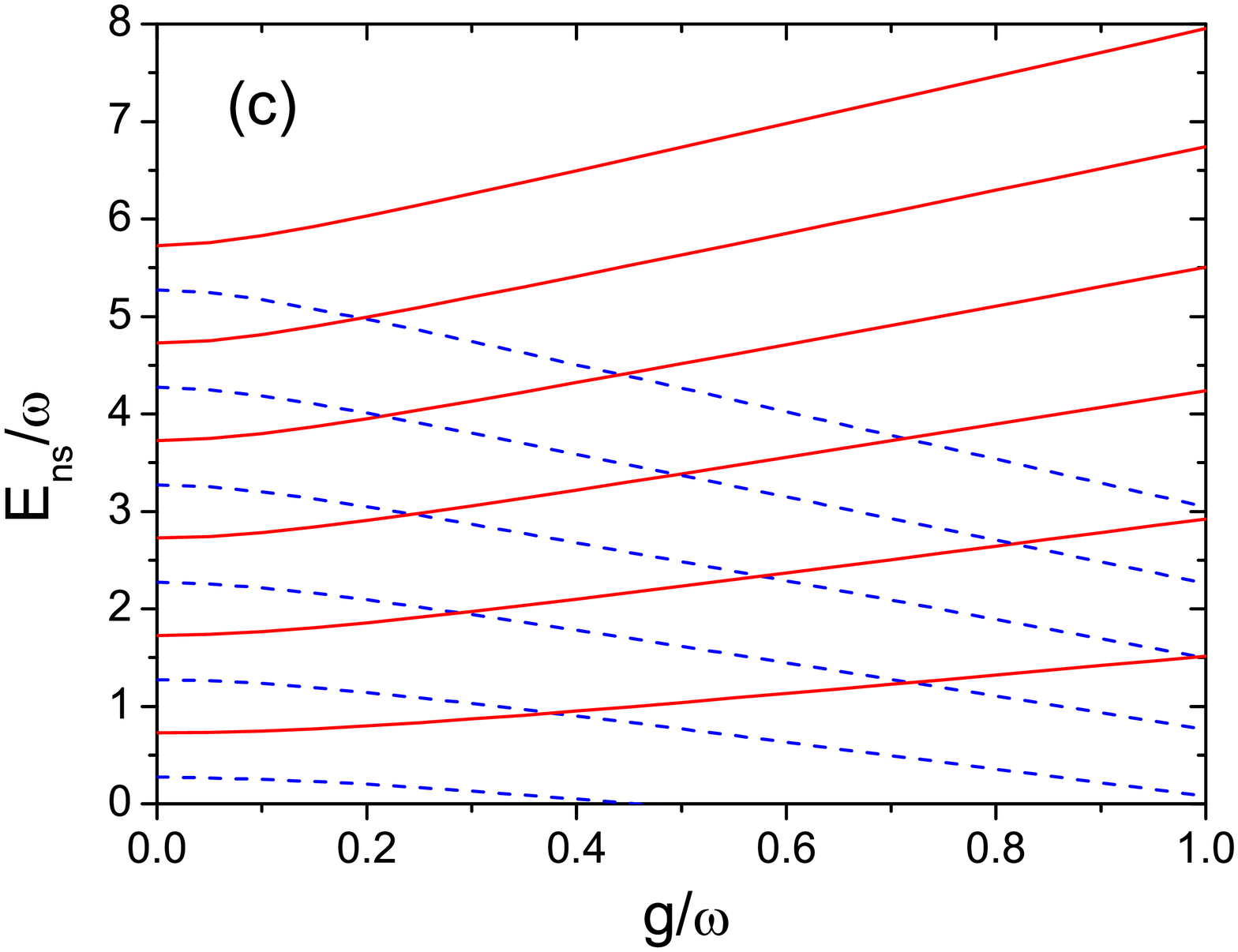}}
\rotatebox[origin=c]{0}{\includegraphics[angle=0,
          height=1.24in]{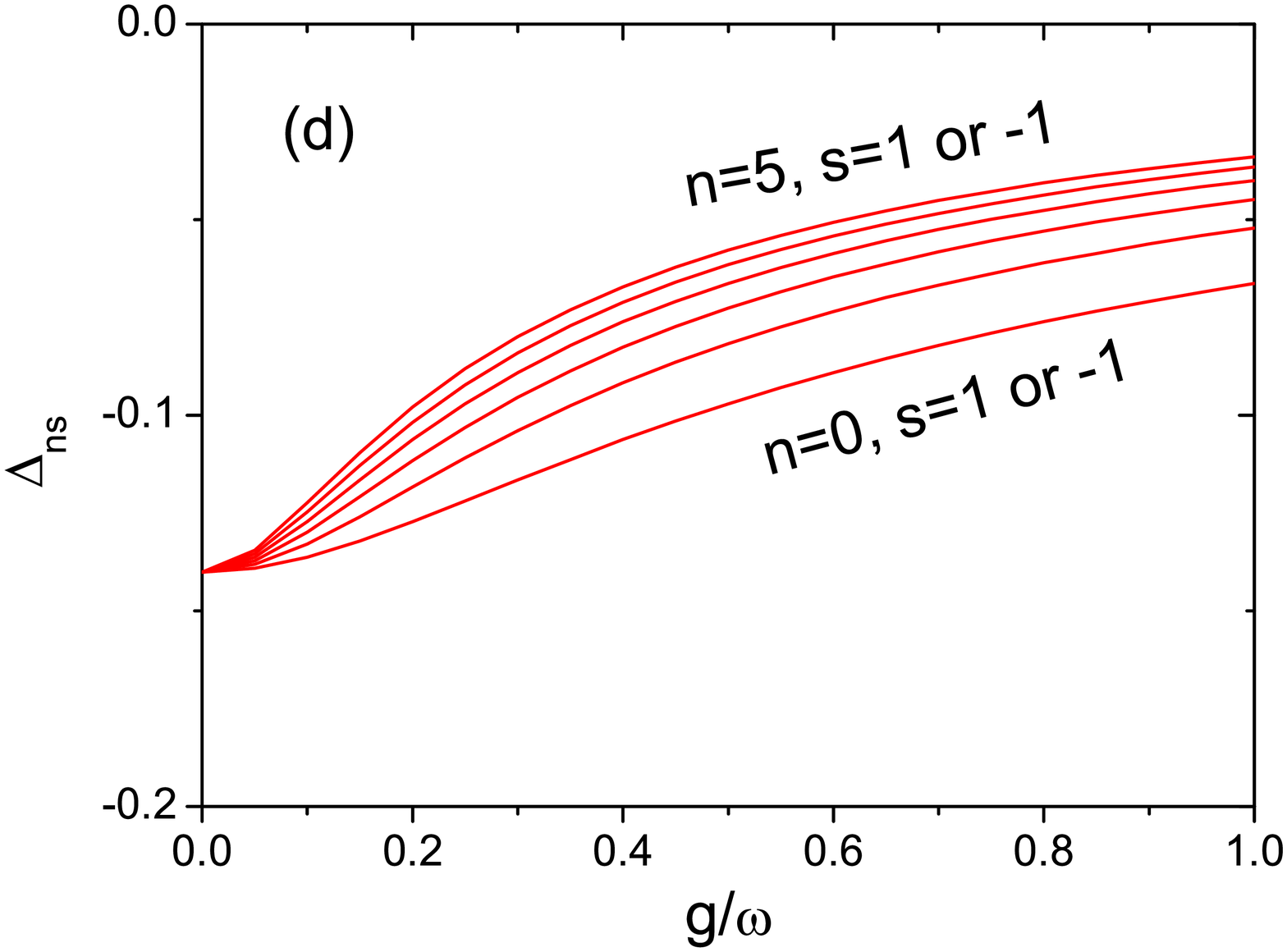}}
\caption {(Color online) The low-lying energy levels of the sub-energy spectrum I and II in unit of $\omega$ as a function of the coupling parameter $g$ when $\lambda=0.7\omega$ and $\epsilon=0.2\omega$, shown in (a) and (c), respectively. The solid lines denote $n=0, 1,\cdots, 5$ and $s=1$ while the dash lines mean $n=1, 2, \cdots, 5$ and $s=-1$. The corresponding $\Delta_{ns}$ are displayed in (b) and (d).}
\end{figure}

For the nth eigenstate with $s$ in the sub-energy spectrum II, from Eq. (43), we have
$$B^{ns}_{n+1}=\frac{(1+\Delta_{ns}^2)(E_{ns}-n\omega)-\lambda(1-\Delta_{ns}^2)+2\epsilon\Delta_{ns}}{g(1-\Delta_{ns}^2)(n+1)}
A^{ns}_n,\eqno{(50)}$$
where $A^{ns}_n$ is a constant to be determined by the normalized condition (29).
The other coefficients $A^{ns}_i$ and $B^{ns}_i$, proportional to $A^{ns}_n$, also satisfy
the same recursion relations (41) and (42) in the sub-energy spectrum I.

In order to compare with the energy spectrum of the Rabi model presented by Braak, here we employ the physical parameters in Ref. [16].
Figs. 5 and 6 exhibit the low-lying energy levels of the sub-energy spectrum I and II as a function of $g$
at $\lambda=0.4\omega$ and $\epsilon=0$ and at $\lambda=0.7\omega$ and $\epsilon=0.2\omega$, respectively.
We can see that the energy spectrum possesses the level crossings between the neighboring eigenstates,
which is dramatically different from that in Ref. [16]. It is expected that such the degeneracies at certain physical parameter values
could produce novel physical phenomena in the two-level system with the light-matter interaction, similar to the two-dimensional electron gas with
spin-orbit interaction under a perpendicular magnetic field [19-21].

\section{IV. Summary}

We have exactly solved the quantum Rabi model (1) in both the occupation number representation and the Bargmann space.
The complete energy spectrum is comprised of two double-fold degenerate sub-energy spectrum I and II.
Such the exact solution can help us to deeply understand the light-matter interaction, especially in strong coupling regimes.
Because the analytical expressions of the eigenvalue $E_{ns}$ in the occupation number representation are completely identical to
those in the Bargmann space, this exact solution for quantum Rabi model is definitely correct.

\section{ACKNOWLEDGEMENTS}

This work was supported by the Sichuan Normal University, the "Thousand
Talents Program" of Sichuan Province, China, the Texas Center
for Superconductivity at the University of Houston,
and the Robert A. Welch Foundation under grant No. E-1146.

\section{APPENDIX}

Braak started from the Rabi model
$$ H_{sb}=\omega a^+a+g\sigma_z(a^++a)+\Delta\sigma_x \eqno{(A1)}$$
in Ref. [16]. After taking the transformations $a\rightarrow \frac{\partial}{\partial z}$
and $a^+\rightarrow z$, then the Hamiltonian (A1) becomes
$$H_{sb}=\left(
\begin{array}{cc}
\omega z\partial_z+g(z+\partial_z)&\Delta\\
\Delta&\omega z\partial_z-g(z+\partial_z)\\
\end{array}\right).\eqno{(A2)}$$
Suppose that $(\phi_1 ~\phi_2)^T$ is the two-component wave function of $H_{sb}$. Then
one has  a coupled system of the first-order differential equations
$$
\begin{array}{rcl}
&(z+g)\frac{d}{dz}\phi_1(z)+ (gz-E)\phi_1(z)+\Delta\phi_2(z)=0,&~~~(A3)\\
&(z-g)\frac{d}{dz}\phi_2(z)-(gz+E)\phi_2(z)+\Delta\phi_1(z)=0,&~~~(A4)\\
\end{array}$$
where $\omega=1$ and $E$ is the corresponding eigenvalue. Braak found that Eqs. (A3) and (A4)
have the following solution
$$\begin{array}{rcl}
~~~&\phi_1(z)=e^{-gz}\sum_{n=0}^{\infty}K_n(x)\Delta\frac{(z+g)^n}{x-n},&~~~~~~~~~~~(A5)\\
~~~~~~~&\phi_2(z)=e^{-gz}\sum_{n=0}^{\infty}K_n(x)(z+g)^n,&~~~~~~~~~~~(A6)
\end{array}$$
where $x=E+g^2$, $E$ can take an arbitrary value, and the constants $K_n(x)$
satisfy the recursive relation (4) in Ref. [16].
Obviously, $\phi_1(z)$ and $\phi_2(z)$ are divergent at $z\rightarrow -\infty$.
Therefore, this two-component solution $(\phi_1 ~\phi_2)^T$ of $H_{sb}$ is
trivial and non-physical due to the divergence of the wave function and the
undetermined eigenvalue.

In order to fix the eigenvalue $E$, Braak employed the unitary transformation
$$U=\frac{1}{\sqrt{2}}\left(
\begin{array}{cc}
1&1\\
T&-T\\
\end{array}\right),\eqno{(A7)}$$
where the operator $T$ satisfies $T(f)(z)=f(-z)$. It is easy to get
$$U^+H_{sb}U=\left(
\begin{array}{cc}
H_+&0\\
0&H_-\\
\end{array}\right).\eqno{(A8)}$$
Here, $H_\pm=\omega z\partial_z+g(z+\partial_z)\pm\Delta T$.
Obviously, $H_-$ can be obtained from $H_+$ by letting $\Delta$ be $-\Delta$.

The time-independent Schrodinger equation for $H_+$ with positive parity reads
$$z\frac{d}{dz}\psi(z)+g(\frac{d}{dz}+z)\psi(z)
=E\psi(z)-\Delta\psi(-z),\eqno{(A9)}$$
which becomes
$$z\frac{d}{dz}\psi(-z)-g(\frac{d}{dz}+z)\psi(-z)
=E\psi(-z)-\Delta\psi(z)\eqno{(A10)}$$
after manipulating $T$ on two sides of Eq. (A9).
Here $\omega=1$ and $E$ is the eigenvalue of $H_+$ (or $H_{sb}$). It is obvious that
$\psi(z)$ and $\psi(-z)=T(\psi)(z)$ in Eq. (A9) or (A10) are correlated due to
the reflection operator $T$.

With the notation $\psi(z)=\phi_1(z)$ and $\psi(-z)=\phi_2(z)$,
Eqs. (A9) and (A10) lead to Eqs. (A3) and (A4), respectively.
Such a notation is the solution of the coupled equations (A3) and (A4)
rather than the single equation (A9) with the presence of $T$.
I note that if and only if
$$G_{+}(x; z)=\phi_2(z)-T\phi_1(z)=\phi_2(z)-\phi_1(-z)\equiv 0 \eqno{(A11)}$$
for any $z$, this notation $\{\psi(z),\psi(-z)\}$ is the solution of Eq. (A9).

Obviously, Braak treated $\psi(z)$ and $\psi(-z)$ as independent wave functions
and neglected the condition (A11).
By requiring the wave function $\{\psi(z),\psi(-z)\}$ to be continuous at $z=0$,
i.e. $G_{+}(x; z=0)=0$, Braak obtained the eigenvalues $E$
(see (3) and Fig. 1 in Ref. [16]). However, the constraint (A11) does not hold for
nonzero $z$ under these eigenvalues $E$ [see the expressions (A5) and (A6)].
So such a wave function $\{\psi(z),\psi(-z)\}$ with a cusp at $z=0$ and the corresponding
eigenvalue $E$ are not these of $H_+$.
Similarly, the solution for $H_-$ with negative parity can be obtained by replacing
$\Delta$ with $-\Delta$.
Therefore, it is out of question that the energy spectrum shown in Figs. 2 and 3 in Ref. [16]
is not that of the Rabi model (A1).

Braak also applied the similar technique to the generalized Hamiltonian by adding
$\epsilon \sigma_z$ to (A1) (i.e. (7) in Ref. [16]). However, the same derivation error
occurs. The energy spectrum depicted in Fig. 4 in Ref. [16] is also incorrect.


\end{document}